\documentclass[sigconf]{acmart}

\PassOptionsToPackage{table}{xcolor}
\newcommand\am[1]{}
\newcommand\cw[1]{}
\newcommand\rd[1]{}
\newcommand\sd[1]{}

\newcommand{\etals}
{et al.'s}

\newcommand{\ie}{{i.e.}}
\newcommand{\eg}{{e.g.}}

\newcommand{\cf}{{cf.}}

\newcommand{\secref}[1]{\hyperref[#1]{Sec.~\ref*{#1}}}
\newcommand{\figref}[1]{\hyperref[#1]{Fig.~\ref*{#1}}}

\definecolor{quoteColor}{HTML}{663399}
\newcommand\pX[1]{{\color{black}\textbf{P#1}}}
\newcommand\pQ[2]{\emph{``#2'' (\textbf{P#1})}}
\newcommand\pQalt[1]{\emph{``#1''}}

\definecolor{takeawayColor}{HTML}{3C8031}
\newcommand\lesson[1]{#1}

\newcommand\surveySystem[1]{\emph{#1}}
\newcommand\designSpaceItem[1]{{\color{teal}\textsc{#1}}} 
\usepackage{xcolor}
\usepackage{soul}
\usepackage{setspace}
\usepackage{multirow}
\AtBeginDocument{%
  }

\copyrightyear{2023}
\acmYear{2023}
\setcopyright{acmlicensed}\acmConference[CHI '23]{Proceedings of the 2023 CHI Conference on Human Factors in Computing Systems}{April 23--28, 2023}{Hamburg, Germany}
\acmBooktitle{Proceedings of the 2023 CHI Conference on Human Factors in Computing Systems (CHI '23), April 23--28, 2023, Hamburg, Germany}
\acmPrice{15.00}
\acmDOI{10.1145/3544548.3580940}
\acmISBN{978-1-4503-9421-5/23/04}

\begin{document}

\title{On the Design of AI-powered Code Assistants for Notebooks}
\author{Andrew McNutt}
\email{mcnutt@uchicago.edu}
\orcid{0000-0001-8255-4258}
\affiliation{%
    \institution{University of Chicago}
    \city{Chicago}
    \state{IL}
    \country{USA}
    \postcode{60637}
}

\author{Chenglong Wang}
\email{chenwang@microsoft.com}
\affiliation{%
    \institution{Microsoft research}
    \city{Redmond}
    \state{WA}
    \country{USA}
}

\author{Rob DeLine}
\email{rob.DeLine@microsoft.com}
\affiliation{%
    \institution{Microsoft research}
    \city{Redmond}
    \state{WA}
    \country{USA}
}

\author{Steven M. Drucker}
\email{sdrucker@microsoft.com}
\affiliation{%
    \institution{Microsoft research}
    \city{Redmond}
    \state{WA}
    \country{USA}
}

\renewcommand{\shortauthors}{McNutt et al.}

\begin{abstract}
    AI-powered code assistants, such as Copilot, are quickly becoming a ubiquitous component of contemporary coding contexts.
    Among these environments, computational notebooks, such as Jupyter, are of particular interest as they provide rich interface affordances that interleave code and output in a manner that allows for both exploratory and presentational work.
    Despite their popularity, little is known about the appropriate design of code assistants in notebooks.
    We investigate the potential of code assistants in computational notebooks by creating a design space (reified from a survey of extant tools) and through an interview-design study (with 15 practicing data scientists).
    Through this work, we identify challenges and opportunities for future systems in this space, such as the value of disambiguation for tasks like data visualization, the potential of tightly scoped domain-specific tools (like linters), and the importance of polite assistants.
\end{abstract}



\maketitle
\section{Introduction}

AI-powered code assistants like GitHub \surveySystem{Copilot}~\cite{copilot} are designed to improve programmers' productivity.
Powered by large language models (LLMs), these tools can automatically generate high-quality code suggestions from a programming context---consisting of both code and natural language, such as comments or docstrings.
Interaction with code assistants typically occurs as part of normal program authoring: the programmer indicates a context in which they would like assistance, which the assistant uses to provide a list of suggestions. The programmer selects a desired code recommendation, adapts it to fit their context, and continues programming.
Through this cycle, users can reduce the burden of boilerplate~\cite{Sarkar22WhatIsItLike}, increase perceived programming productivity~\cite{ziegler2022productivity, vaithilingam2022expectation}, and receive guidance on how to address unfamiliar tasks~\cite{barke2022grounded}.

Given this facility to improve developer experience,
these AI-powered code assistants have the potential for enormous impact in \emph{computational notebooks}, such as Jupyter~\cite{perez2007ipython}.
However, the simple text-based interactions suited to more traditional code editors do not naturally fit into notebook workflows.
For instance, in notebooks, programming is not limited to textual inputs but can involve a wide variety of multi-modal data---such as code, markdown, data tables, and plots.
Notebook users often write small, loosely structured code snippets, which they execute interactively to understand the code and data~\cite{kery2018story}.
To support this often exploratory style, these code snippets are often written and executed in a nonlinear order, with parallel solutions being examined in an iterated and interleaved manner~\cite{weinman2021fork}.

Without accounting for these differences, a purely code-based interface to code assistants would make it challenging for notebook users to specify the desired context, understand assistant suggestions, and adapt them into their work.
In this work, we seek to enable future system designers by considering

\begin{enumerate}
    \item \emph{What choices are available in the design of AI-powered code assistants in notebooks?}
    \item \emph{What do users expect from such assistants in this context?}
\end{enumerate}

To answer these questions, we conducted two studies, a design space analysis, and a semi-structured design study.

In the first of these, we sought to characterize the design space by surveying interaction designs in notebooks whose goal is to improve the end-user programming experience through code generation.
Despite their often ad hoc or domain- and algorithm-specific design, these systems---which range from integrating graphically specified elements~\cite{kery2020mage} to live spreadsheet manipulations~\cite{mito}---provide valuable insights for the design of AI-powered assistants more generally.
We extrapolate these design choices into a collection of design concerns (\secref{sec:design-space}), which we display in
\figref{fig:designs-space-a} and \figref{fig:designs-space-b}.
We classify the design space based on interface components (user gestures, model artifacts, disambiguation, and refinement interfaces) as well as the relationship between these components (code context, provenance management, model specialization, and customizability).
We phrase our concerns in a generalizable manner that can be used to generatively explore  the space.

Building on the structure of this space, we conducted a second study that sought to understand data scientists' expectations of AI-powered code assistants in notebooks.
In this study, we interviewed 15 professional data scientists about their preferences for tools in this space.
We presented participants with various designs that probed their perceptions of context specification, suggestion disambiguation, result adaptation, and code provenance situated within several data analysis and visualization tasks (\secref{sec:interview}).

While different participants preferred different options in the design space, they were unanimously enthusiastic about the potential that our probes suggested.
Our study revealed a rich set of predilections about this style of system (\secref{sec:analysis}), which we summarize in \autoref{tab:takeaways}.
Most notable among these:
the importance of polite interfaces that respect users' agency and flow, but are sufficiently prominent to promote usage;
the potential that linter-like assistants that highlight inappropriate usage in an approachable manner might have when scoped to a specific medium or task (such as data science or notebooks);
that assistants might usefully take on a variety of interface forms throughout the notebook to aid in a corresponding array of tasks.

Integration of code assistants backed by AIs offers rich potential to radically improve notebook users' programming experience and efficacy.
Through our characterization of the design space and elicitation of realistic-user expectations and opinions, we hope to empower future designers to build more helpful code assistants.

\section{Related Work}

Our work is informed by prior work on code generation models, interfaces for interacting with those models, design interventions in notebooks, and code suggestions more generally.

\paragraph{Code Generation}
Program synthesizers (which we refer to more generally as code generation) seek to reduce programming effort by automating challenging or repetitive programming tasks.
These techniques automatically generate code based on high-level specifications---such as through demonstrations, input-output examples, natural language, and partial implementations~\cite{gulwani2017program}.

Program synthesizers come in a variety of styles, but of particular interest to our work are code generation models, especially those powered by LLMs.
These models automatically learn program concepts from large-scale code corpora and  are typically not limited to a specific language or a task domain.
For example, Codex~\cite{chen2021evaluating} (the model used in \surveySystem{Copilot}), InCoder~\cite{fried2022incoder}, and   CodeGen~\cite{nijkamp2022conversational} are GPT-style code generation models~\cite{gptOpenai} trained on GitHub repositories (among other sources) that support code generation from texts and partial implementations in most mainstream programming languages.
A number of code assistants have been developed based on these models~\cite{amazonCodeWhiperer, fauxpilot, tabnine, copilot}.

Despite their support for general languages and tasks, these models are no panacea: in addition to requiring significant natural resources to train~\cite{bender2021dangers}, they often handle specific application scenarios poorly.
These issues can sometimes be addressed through prompt engineering~\cite{reynolds2021prompt,jain2021jigsaw} (rephrasing the specification in languages that the model understands) or fine-tuning~\cite{Cobbe21Training, chen2021evaluating} (additional training on datasets that represent the application domain).

Our work focuses on \surveySystem{Copilot}, not because we seek to develop a better understanding of it in particular, but instead using it as a representative of a wider class of LLM-driven code assistants.

\paragraph{Interacting with Code Assistants}
A major open question is how to employ code generation models in a way that is understandable and maintains users' sense of control. This has long been a concern of mixed-initiative systems~\cite{horvitz1999principles}, but only recently have code assistants become powerful enough to enable study of their interfaces.

A variety of interface strategies have been developed to adapt code assistants to particular domains.
\citet{jiang2022discovering, jiang2021genline} describe GenLine: an IDE-based LLM-powered code assistant for HTML/JS  applications. Users guide synthesis through natural language prompts and control context through ``code fences'', which explicitly indicate what is and is not relevant to code generation.
\citet{ferdowsifard2020small} provide a live programming interface that involves an always-on display of values in small python programs that use local test cases to guide synthesis.
To aid users in providing better specifications (or prompts) to the code generation models, AI Chains~\cite{wu2022ai} and PromptIDE~\cite{Strobelt22Interactive} help users experiment and refine their prompts.
Our work draws on these design patterns to understand the space of possible interface elements that might be used in our domain.
In addition to these tools, some systems focus specifically on code generation in notebooks (\secref{sec:design-space}).

GitHub \surveySystem{Copilot}~\cite{copilot} is the first LLM-based code assistant powered to reach widespread usage.
It provides a text-based interface where users provide tacit descriptions of context, either through description or partial programs.
Recent studies of user experiences have sought to characterize the user experience and perceptions of interacting with AI-powered assistants.
\citet{barke2022grounded} find that users tend to use \surveySystem{Copilot} either in an acceleration mode (in which it is used to enhance their efficiency) or an exploration mode (in which users are aided in understanding the available program space).
\citet{Sarkar22WhatIsItLike} find that \surveySystem{Copilot} usage has similarities to search, compilation, and pair programming.
\citet{Jayagopal22Exploring} analyze how novices learn to use code generation tools and highlight the tensions between triggered and triggerless initiation and communication of code generation.
Other studies suggest that code generation quality depends on user experience level~\cite{ziegler2022productivity}, that less experienced users had difficulty understanding, editing, and debugging generated code~\cite{vaithilingam2022expectation}, and that users may not trust generated code compared to code produced by a human pair programmer~\cite{imai2022github}.
Our work extends these efforts by considering AI-powered code assistants in a particular medium.

\paragraph{Intervening in Computational Notebooks}
Computational notebooks, like Jupyter~\cite{perez2007ipython}, have become a popular environment for data scientists to conduct some aspects of their work~\cite{lau2008pbd, rule2018exploration}. While AI-powered code assistants like \surveySystem{Copilot} enhance software development environments, researchers meanwhile have explored other types of code assistance in computational notebooks~\cite{weinman2021fork, merino2020bacat}.
For instance, \citet{kery2017variolite} explore design interventions for adapting the notion of history to the specifics of notebooks, while \citet{head2019managing} apply program slicing as means to automatically organize notebooks.
To help with end-user programming in notebooks, tools such as \surveySystem{Mage}~\cite{kery2020mage} let users specify code using graphical elements (\eg{} through a visualization widget). \surveySystem{Mito}~\cite{mito} and \surveySystem{B2}~\cite{wu2020b2} allow users to manipulate  data directly through spreadsheet interfaces and dashboards, respectively, where the interactions are recorded as code.
Additional notebooks have been developed to address notebook usability issues~\cite{lau2020design}.
For instance, \surveySystem{Glinda}~\cite{deline2021glinda} is a notebook with live programming and declarative language support, while Lodestar~\cite{Raghunandan22Lodestar} enables automation of data science workflows through analysis templates provided via automated recommendation.
Our efforts are closely related to these, as we seek to enhance the end-user experience of notebooks by better understanding the design of code assistants in notebooks.

\begin{figure}[t]
    \centering

    \includegraphics[width=\linewidth]{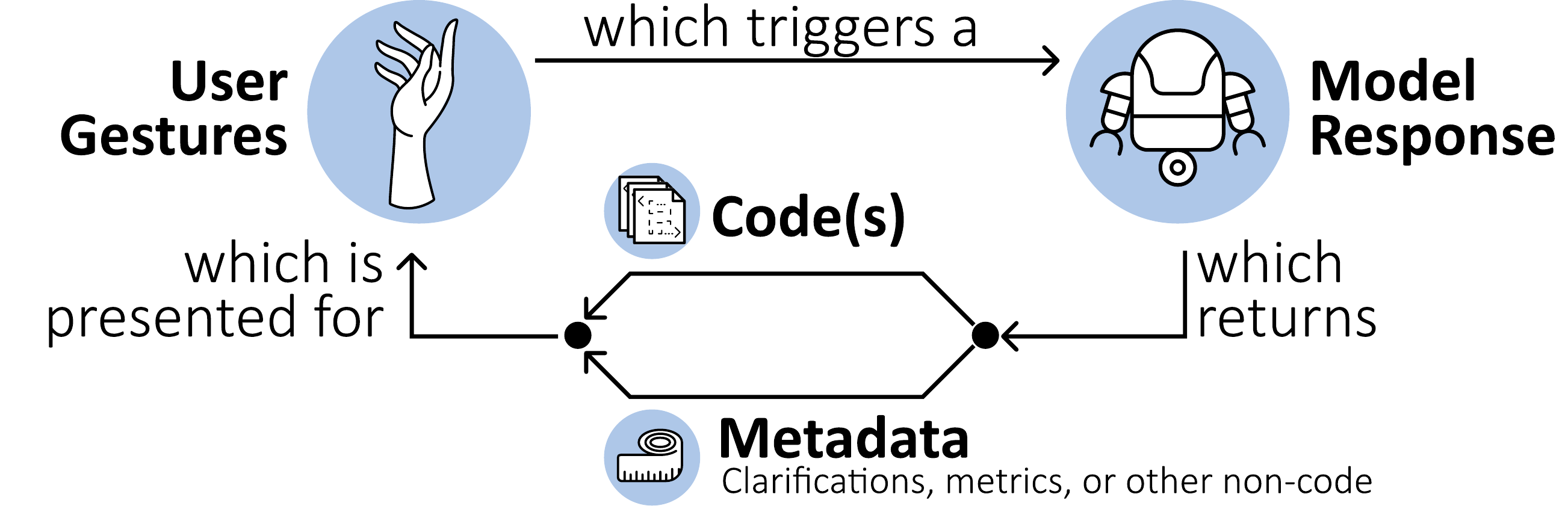}

    \caption{The \emph{specification-refinement loop} interaction underlying our design space.
    }
    \label{fig:interaction-model}
    \Description{A flow chart of actions in the loop. The start position is labeled "user gestures", which has an arrow labeled "which triggers a" that is connected to a node labeled "Model Response". This node has a single outgoing arrow labeled "which returns", which leads to nodes labeled "Code(s)" and "Metadata". These share an arrow labeled "which is presented for" that leads back to the start. In effect, this causes the whole illustration to read ‘User gestures which trigger a Model Response which returns "Code(s)" and or "Metadata" which is presented for "User Gestures"’.}
    \vspace{-2em}
\end{figure} 
\paragraph{Code Suggestions}
A common strategy for making code easier to produce is through in-editor code suggestions, of which generative assistants (such as \surveySystem{Copilot}) are only the latest incarnation. This approach is ubiquitous (arising as early as 1985~\cite{HistoryOfCodeCompletion}), appearing in IDEs, purpose-specific editors like Jupyter~\cite{merino2020bacat} or Excel~\cite{excelAutcomplete}, as well as specific-domains
such as exploratory data analysis~\cite{li2021edassistant}, data visualization~\cite{rong2016codemend}, and computational notebooks~\cite{li2021nbsearch}.
This strategy might take the visual form of an autocomplete (as in IntelliSense~\cite{MicrosoftIntelliSense} or Calcite~\cite{mooty2010calcite}) or through UIs with search-like signifiers like a query bar (as in Blueprint~\cite{brandt2010example} or Bing Developer Assistant\cite{zhang2016bing}).
The interface for asking for a suggestion (and therein specifying intent and context) can involve caret position~\cite{mooty2010calcite, omar2012active, brandt2010example, MicrosoftIntelliSense}, code comments~\cite{ye2005reuse}, partial implementations~\cite{reiss2014seeking}, integrated examples~\cite{oney2012codelets}, or explicit querying.
\citet{liu2021opportunities} find that code search can use a variety of input modalities, including free text, source code, API descriptions, input-output examples, test cases, and UI sketches---typically without direct control from the user~\cite{grazia2022code}.
Some works~\cite{rong2016codemend, brandt2010example} highlight the value of integrating documentation, provenance, or justification as part of the generated results.
In modeling interactions with generative code assistants (\secref{sec:design-space}), we did not analyze purely search-based interfaces, although generation and search are typically seen as closely-related interactions~\cite{Sarkar22WhatIsItLike} (\secref{sec:search}).
For instance, \citet{xu2022ide} compare search and generation through an assistant that mixes both approaches, finding that each modality is beneficial for different tasks.
Our studies draw on the patterns found in these interaction forms, however, our exclusion of search systems is a limitation of our approach.

\section{Design Space}\label{sec:design-space}

To explore design options for AI-powered code assistants, we first analyzed the broader design space of how code assistants are deployed in computational notebooks. We consider systems that emit textual code based on some user gesture\footnote{
    We use the term \textit{gesture} to cover a variety of input modalities, such as direct manipulation, form usage, coding, and natural language.
} to assist user programming, with or without AI backends.
Through this work, we develop a characterization of this design space that describes both \emph{interfaces} and \emph{interactions} available in such systems.
Our approach draws on observational and reflective approaches~\cite{Munzner22DesignSpaces}.

From the observational side, we sought to assemble a maximum variation sample~\cite{patton1990qualitative} of this interface form to understand as many different approaches as available. We did so by searching Google Scholar for systems involving notebooks and code generation from which we iteratively snowball sampled. Through this process, we identified 14 systems:
\surveySystem{B2}~\cite{wu2020b2},
\surveySystem{bamboolib}~\cite{bamboolib},
\surveySystem{Copilot}~\cite{copilot},
\surveySystem{Gauss}~\cite{bavishi2021gauss},
\surveySystem{Glinda}~\cite{deline2021glinda},
\surveySystem{Hex}~\cite{hex},
\surveySystem{Jigsaw}~\cite{jain2021jigsaw},
\surveySystem{Lux}~\cite{lee2021lux},
\surveySystem{Mage}~\cite{kery2020mage},
\surveySystem{Mito}~\cite{mito},
\surveySystem{Observable data table}~\cite{ObservableDataTable},
\surveySystem{Tabnine}~\cite{tabnine},
\surveySystem{VizSmith}~\cite{bavishi2021vizsmith},
and \surveySystem{Wrex}~\cite{drosos2020wrex}.
These systems range from domain-specific code assistants (\eg{} assistants for spreadsheet data manipulation or statistical analysis) to live programming environments and frameworks for building GUI-embedded widgets.
While there are other systems of interest, we believe this selection is sufficient for our analysis.
We provide example images of each system in the appendix.
Despite the diversity of these systems, some corners of this relatively new space remain under-explored, and some dimensions may simply reflect the design choices that Jupyter or its Extension API impose.
To address these biases, we augmented our observations with a series of identified properties.
For instance, in our analysis of where code assistants are located within the notebook, we observed systems using only \designSpaceItem{Ambient}, \designSpaceItem{Inline}, and \designSpaceItem{Application} style assistants but missing \designSpaceItem{cell}-level designs, so we extended the design space to include this option.
Finally, while closely connected, we exclude search-based systems to limit scope and because generative assistants seem~\cite{Sarkar22WhatIsItLike} to be used differently than search systems.

\subsection{The Design Space}

\begin{figure*}
    \centering
    \includegraphics[width=0.9\linewidth]{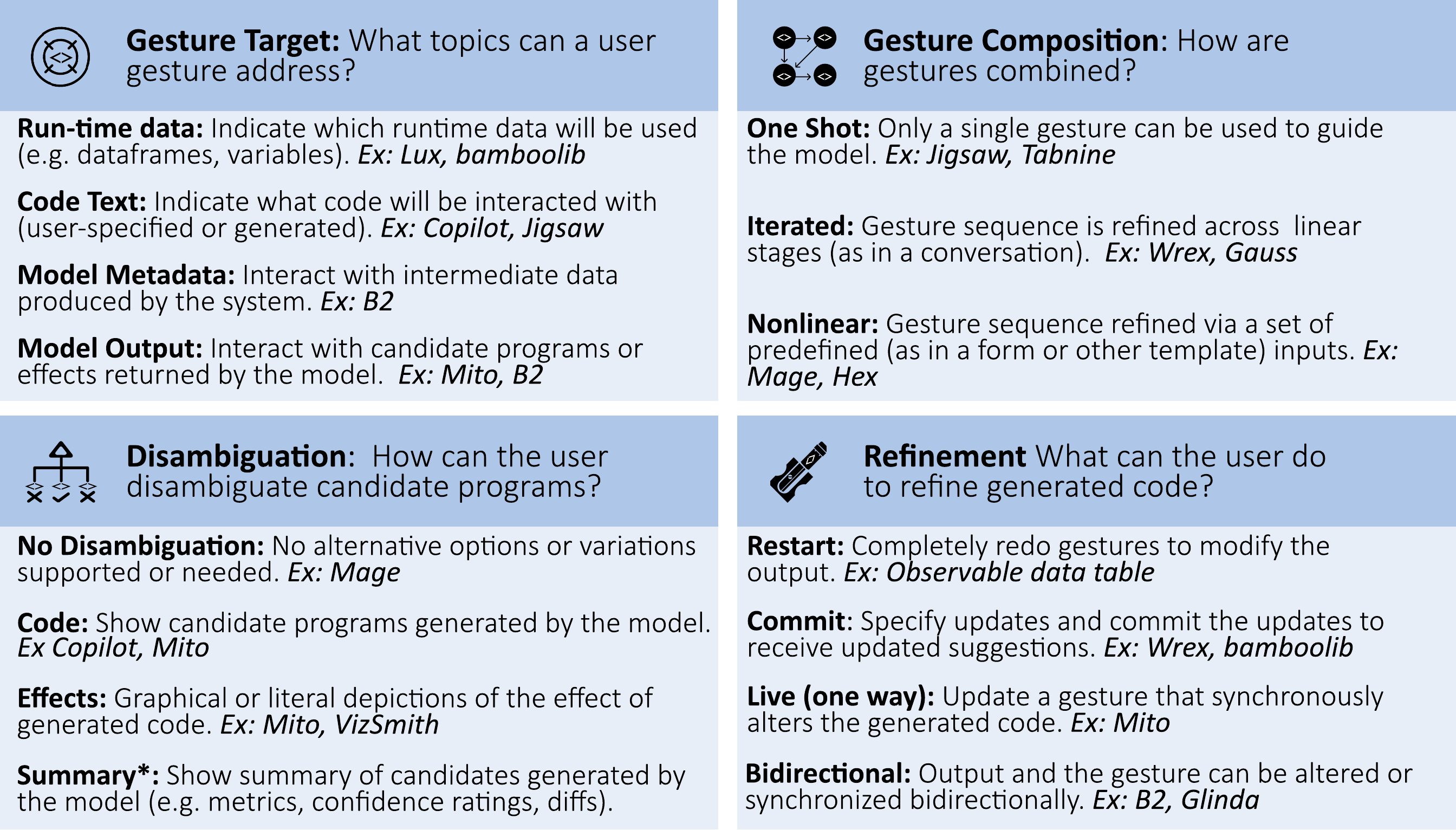}
    \caption{
        The components of our design space considering interactions with the code assistant. Among these topics, our interview study investigates \emph{Disambiguation} interaction most directly.
    }
    \label{fig:designs-space-a}
    \Description{An set of four boxes showing the first half of the design space. It is divided into four boxes arranged in a grid.
    }
\end{figure*}

\begin{figure*}
    \centering
    \includegraphics[width=0.9\linewidth]{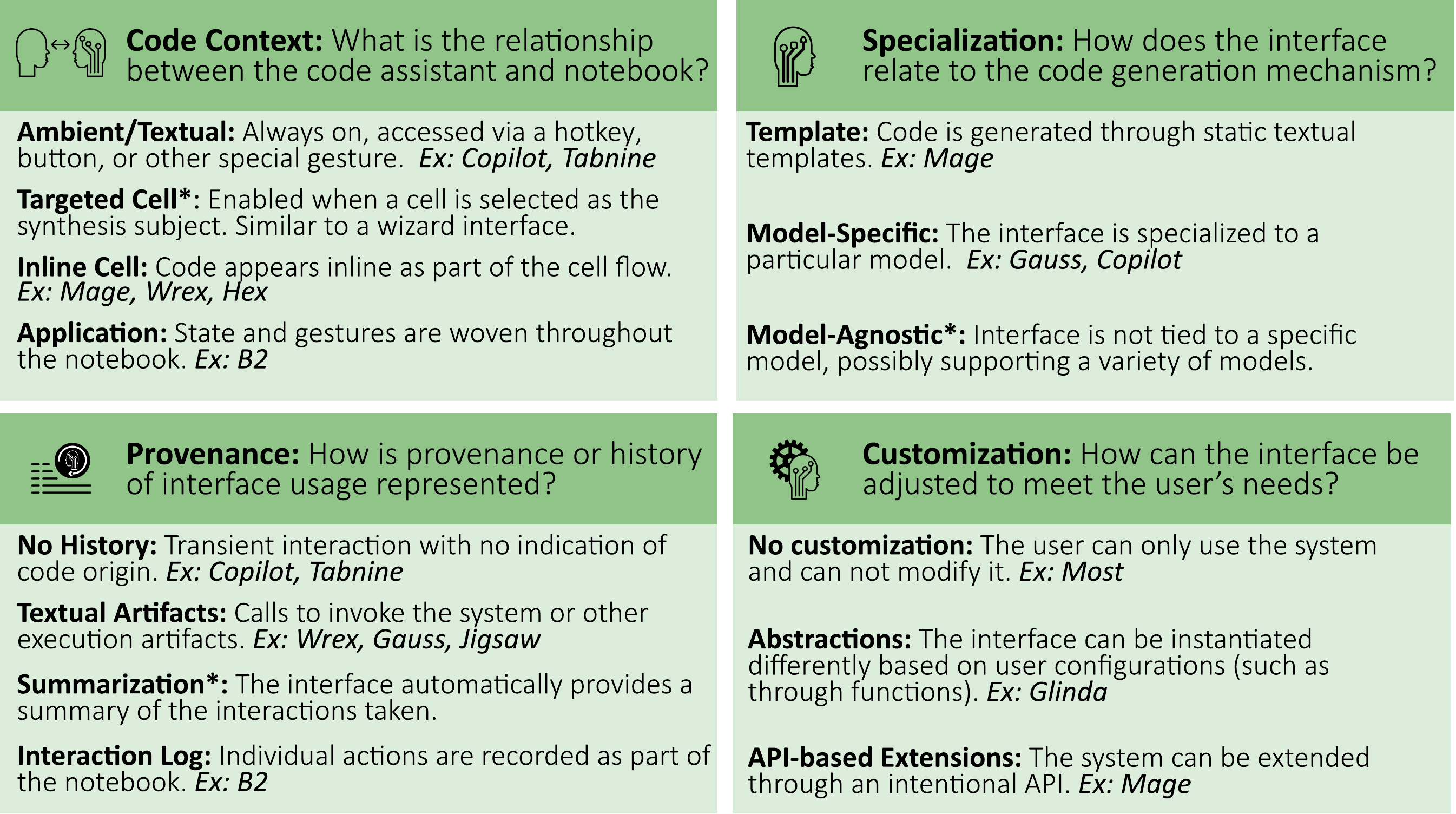}

    \caption{
        The parts of our design space considering the relationship between code assistant and other system components mediated by the interface. Our interview study considers the \emph{Code Context} and \emph{Provenance} interface relationships.
    }
    \label{fig:designs-space-b}
    \Description{A set of four boxes showing the second half of the design space. It is divided into four boxes arranged in a grid.
    }
\end{figure*}

We describe the interaction between the user and the code assistant as a \emph{specification-refinement loop} (\figref{fig:interaction-model}). In this loop, the user specifies the programming goal using a sequence of
gestures, then the system returns code or metadata to the user, which forms the basis for the next cycle of interaction.
We capture how different systems embody this loop along two principal categories, namely interactions with the assistant (\secref{sec:interactions}) and relationships between the code assistants and other notebook components (\secref{sec:interface}), which is based on Beaudouin-Lafon's dichotomization of interactions versus interfaces~\cite{beaudouin2004designing}.
Within these categories, the reviewed systems vary along several dimensions.
These categories and their dimensions constitute our design space
(\figref{fig:designs-space-a}, \figref{fig:designs-space-b})
wherein systems can be described by choosing a value (or values) from each dimension.
The remainder of this section explains each dimension.

\subsubsection{Interactions}
\label{sec:interactions}
We begin by looking at the dimensions that describe the user interactions with the system.

\paragraph{Gesture Target} Each user gesture targets an entity.
In our context, these include components from the notebook environment and artifacts produced by interaction with the assistant.
Notebook targets include \designSpaceItem{Runtime Data}, such as variables and data frames from the execution context (as in \surveySystem{bamboolib}~\cite{bamboolib}),  and \designSpaceItem{Code Text}, such as text-represented program snippets in code cells (as in \surveySystem{Jigsaw}~\cite{jain2021jigsaw}).
The model-generated targets include \designSpaceItem{Model Metadata}, such as confidence in textual predictions, model parameters or partial programs (as in \surveySystem{B2}~\cite{wu2020b2}), and
\designSpaceItem{Model Output} such as the code generated by the model or visualizations of its effects (as in \surveySystem{Mito}~\cite{mito}).
This rich space of gesture targets defines the basic means through which the user can communicate with the code assistant.

\paragraph{Gesture Composition}
It is often necessary to sequence or compose different user gestures to express more nuanced intent.
The simplest way is \designSpaceItem{One Shot} interaction, wherein each user gesture triggers an independent interaction with the assistant, as is the case with \surveySystem{Tabnine}~\cite{tabnine}.
For multi-step interactions between user and system, systems can employ either \designSpaceItem{Iterated}, in which  gestures are used to iteratively or conversationally refine intent (as in \surveySystem{Wrex}~\cite{drosos2020wrex}),  or \designSpaceItem{Nonlinear}, in which gestures are inputted in an unprescribed order like in a template---as in the \surveySystem{Hex}~\cite{hex} query builder.
The means of composition informs how state is handled and how the UI should present interactions with that state.

\paragraph{Disambiguation} Due to incompleteness of the user specification and uncertainty of the underlying model, many systems produce a set of candidate suggestions rather than a single result.
We highlight the presentation of these possible results as a potentially distinct step from the usage of the final results---although they may be colocated (as in \surveySystem{Copilot}).
We observed several approaches to this task.
Many systems have \designSpaceItem{No Disambiguation} mechanism and only present a single suggestion to the user (as in \surveySystem{Mage}~\cite{kery2020mage}).
Others show multiple candidate \designSpaceItem{Code} options, as in \surveySystem{Copilot's} alternate selection tab.
Some preview the \designSpaceItem{Effects} of running the code, as in \surveySystem{VizSmith's}~\cite{bavishi2021vizsmith} multiple chart options.
Our model also allows \designSpaceItem{Summary} information, which includes designs like surfacing token confidence~\cite{weisz2021perfection} and textual~\cite{barke2022grounded} or semantic~\cite{wang2022diff} diffs.

\paragraph{Refinement} The user may need to refine or adjust their specification after the initial gesture. For instance, a particular recommendation may necessitate that the user adjust their mental model of what can be expected from the assistant (as in the user-synthesizer gap~\cite{ferdowsifard2020small}) and then update their prompt accordingly.
Support for refinement defines to what degree the user can modify their initial specification.
The most basic support is \designSpaceItem{Restart}, wherein the user needs to redo gestures to update their specification (as in \surveySystem{Observable Data Table}~\cite{ObservableDataTable}).
Some systems support a \designSpaceItem{Commit} action, in which the user can specify updates to the specification while the system holds previous gestures in state, and then the suggestion is updated after commit, perhaps via an ``update'' button---such as updating form fields in \surveySystem{Wrex}~\cite{drosos2020wrex}.
Other systems employ synchronous or \designSpaceItem{Live (One Way)} updates, in which the suggestion is updated immediately after the gesture, such as how \surveySystem{Mito}~\cite{mito} updates the generated code immediately after interactions with its spreadsheet.
Some systems, such as \surveySystem{Glinda}~\cite{deline2021glinda} and \surveySystem{B2}~\cite{wu2020b2}, extend this idea through \designSpaceItem{Bidirectional} updates~\cite{hempel2022maniposynth, hempel2019sketch, williams2021provenance}, in which modifications to the output update the gesture input as well.
While bidirectional updates are ideal---as they maintain consistency between specification and model state---this approach is rare in practice due to the high complexity of synchronizing model, text, and output updates.

\subsubsection{Interface Relationships}
\label{sec:interface}
We next consider the parts of our space that describe the relationship between the code assistant interface and other system components.
\paragraph{Code Context}
A key design choice for the code assistant is its relative location within the notebook UI. This location, in turn, suggests the scope of information available to the code assistant. We list these from the most localized context to the most global. In \designSpaceItem{Ambient/Textual} interfaces, the code assistant follows the text cursor, and the assistant is invoked via a special action such as a hotkey, as in \surveySystem{Tabnine}~\cite{tabnine}.
In \designSpaceItem{Targeted Cell} interfaces, the code assistant apparently resides within notebook cells to support editing or code generation directly within a given cell, as in \figref{fig:context-models-wizard}.
Code assistants presented as \designSpaceItem{Inline Cells} appear as a contiguous part of the notebook flow, sometimes as a distinct cell type of its own, as in \surveySystem{Mage}~\cite{kery2020mage} or \figref{fig:context-models-inline}. \designSpaceItem{Application}-style interfaces appear integrated into the notebook at a level above the cells, such as through a side panel or other higher-level control, as in \surveySystem{B2}~\cite{wu2020b2}  or \figref{fig:context-models-sidepanel}.
The location of the assistant interface informs what information the user can provide to the system for code generation.
\paragraph{Specialization}
Behind any code assistant is a mechanism for transforming inputs into code---such as an ML model, a search engine, an algorithm, or other heuristics---however, there are trade-offs between mechanism exploitation and the interface flexibility  to accommodate different input modalities.
For instance, in \designSpaceItem{Model-Specific} interfaces, the assistant is customized to assist the model, such as in the drag-and-drop interaction used to construct computation traces in \surveySystem{Gauss}~\cite{bavishi2021gauss}.
While \designSpaceItem{Template}-based interfaces are constructed from simple static or textual templates (as in \surveySystem{Mage}~\cite{kery2020mage}), which can enable a much wider array of input styles.
Finally, \designSpaceItem{Model-Agnostic} assistants support a variety of different model types (perhaps even exposing that functionality to the end-user), however, this can come at the cost of more restricted input modalities, such as the text or tabular data input common to many models.
\paragraph{Provenance}
The history of code generated by the assistant---such as what interactions led to the final generated code---can serve as the documentation for auditing, analysis, and sharing.
Many systems provide \designSpaceItem{No History} of interactions with the assistant, as \surveySystem{Copilot}~\cite{copilot}.
Some systems leave \designSpaceItem{Textual Artifacts} of their usage, such as the library calls used to initiate \surveySystem{Jigsaw}.
Others included an explicit \designSpaceItem{Interaction Log}, such as \surveySystem{B2}'s~\cite{wu2020b2} replayable history.
Automatic \designSpaceItem{Summarization} of interactions can be used to describe the usage from a high level.
The interface for displaying provenance may be contained in cells or may be baked into other parts of the interface---such that interactions are privately held in local state.
\paragraph{Customization}
Enabling end-users to alter the code assistant interface allows the system to adapt to different users and tasks.
Yet, most code assistants we examined had \designSpaceItem{No Customization} support.
Customization can be achieved through \designSpaceItem{Abstractions} where the system provides modifiable abstractions (akin to functions) for end-users to customize the system, as in \surveySystem{Glinda}~\cite{deline2021glinda}.
Similarly, \designSpaceItem{API-based Extensions} allow the user to configure properties of the interface through provided APIs, as in \surveySystem{Mage}~\cite{kery2020mage}.
Creating tools with extensibility in mind may help the end-user avoid limitations but may complicate interface design.

\bigskip
\noindent
Besides the above design considerations, there are other elements that vary among existing systems but do not necessarily alter the user experience.
These include elements such as
whether there is a secondary notation~\cite{blackwell2003notational} integrated into a code assistant or the sorts of input modalities  available to the user.
Because notebooks and assistants are an active area of research and design~\cite{lau2020design}, we forgo classification of such features rather than engaging in the quixotic task of enumerating all possible designs of such elements, instead seeking to provide a high-level description of the types of patterns that can occur in their intersection.
While the social roles that code plays (such as being proxies for trusted colleagues) do affect the types of features users perceive as being useful, we forgo modeling the social context surrounding these assistants.
Such social interaction with and through code is its own rich topic of study---even just within data science~\cite{kim2017data, epperson2022strategies, crisan2020passing}---and extends beyond the scope of this work.
However, understanding the role that code assistants can play in those relationships is valuable future work.

\subsection{Model validation}

We validated our design space, by reflecting on its relationship to other systems and consulting experts on \surveySystem{Copilot}-style interfaces.

Code assistants are a relatively new form of interface, and innovation will continue.
We have intentionally tried to construct our space in such a way that future designs may be located within it, but may not be predicted by it.
That is, following Beaudouin-Lafon's \cite{beaudouin2004designing} typification of design spaces, ours seeks to be descriptive (characterizing what has been done) and generative (prompting subsequent designs).
We do not seek to be evaluative in our construction both because the space is evolving and so best practices may change with time, as well as because significant prior work has explored the role of evaluation in coordination with AI models (\cf{} \citet{amershi2019guidelines}).
Other design spaces might be reasonably formed to describe this space given a different set of priorities.
The analysis available through this space is one relating to the identification of similarities between systems and, potentially, the missing elements within a given system. For instance, FlashFill \cite{gulwani2011automating} can be described as an \designSpaceItem{Ambient}, \designSpaceItem{Model-Specific} interface focused on \designSpaceItem{Code Text} in a \designSpaceItem{Nonlinear} manner. As in most~\cite{williams2021provenance} spreadsheets, it has \designSpaceItem{Live (one way)} refinement. It provides \designSpaceItem{No History} and allows \designSpaceItem{No Customization}. Further, this system has  \designSpaceItem{No Disambiguation}, an absence addressed in follow-up works~\cite{mayer2015user, narita2021data}.
Despite this comparison, we only claim that this analysis covers the case of notebooks.
Our design space is meant to dovetail with related analyses, such as Jayagopal \etals{}~\cite{Jayagopal22Exploring} notions of triggered vs triggerless initiation and result communication in code generators.

We presented our design space to several researchers with expertise with \surveySystem{Copilot}, who had run user studies themselves on the topic. These experts received the design space favorably and spoke positively about its value for the design of future systems in this genre. While expert opinion is inherently subjective, their review provides a coarse means of validation. We continue to explore aspects of our design space through our interview study (\secref{sec:interview}).

\section{Interview study}
\label{sec:interview}

To investigate real-world perceptions of elements of our design space we conducted a semi-structured interview study. As the comparison of all design combinations is prohibitively expensive,  we focus our study on four under-explored elements that complement prior studies~\cite{barke2022grounded, jiang2021genline, weisz2021perfection, Sarkar22WhatIsItLike}. In particular, we focus on code context,  disambiguation techniques, adaptation methods, and designs for provenance exploration. %
While we did not specifically construct designs relating to other aspects of our design space (\eg{}  expectations of  customization), these topics arose organically in interviews.

\subsection{Methodology}

We recruited 15 participants (denoted \pX{1} - \pX{15}) from a large data-driven software company.
Participants were contacted based on job title (including roles such as \emph{data scientist}) through the company address book.
We invited those who self-identified on an intake survey as having experience with notebooks and \surveySystem{Copilot} (or another AI-based code assistant) to participate in our interviews.
Despite their response to the intake form,
\pX{3}, \pX{9}, and \pX{15}
did not have experience using an AI-powered code assistant. We include them in the study because their opinions, inclinations, and skepticism about such tools are also informative.
Participants were compensated with a \$50 USD Amazon gift certificate.

Participants reported a median of 1--3 months of experience with an AI-based assistant, 1--2 years in their current roles, 3--5 years of experience using notebooks, and more than 5 years of doing data science work. All participants were 25--34 years old and had at least master's degrees in a CS-related field.
Further demographic data was not collected.
The semi-structured interviews, conducted via video conferencing, alternated between questions regarding participant experiences and presentation of design probes (Figs. 4-8).
We used slide-based prototypes instead of developing interactive prototypes to avoid low-level feedback (\eg{} implementation bugs or minor elements of visual design), following the methodology used by \citet{kery2020mage}.
We did not collect quantitative scores on the presented designs because the semi-structured format of the interview caused some topics to be considered more deeply than others.
Interviews lasted an average of $73\pm10$ minutes.

We recorded and transcribed the interviews. The first author open-coded the transcriptions and built a set of tentative themes, which the research team periodically met to discuss, critique, and iterate on until saturation was reached.

\subsection{Study design}

We next present the issues considered in our study
and the various designs used to explore them.
Discussion of these topics was guided by simple data analysis and visualization scenarios, such as merging two datasets. The details of these settings did not substantially affect discussion, so we forgo description here.
The full study instrument is in the supplement.

\paragraph{Code Context}
Each code generation model has its own notion of context that informs what information the model requires to generate code.
For instance, in \surveySystem{Copilot}, context consists of a fixed length of text around the text caret.
Complicating this specification is that a user's understanding of code context may differ from the assistants, potentially yielding a mismatch between the provided and expected suggestions.
Through consideration of this topic, we sought to understand: \emph{what are users' perceived needs for context specification}?
In our interviews, we considered three design options a \designSpaceItem{Targeted cell}-style option (\figref{fig:context-models-wizard}'s Wizard), a between-cell or \designSpaceItem{Inline}-style option (as in \figref{fig:context-models-inline}), and a notebook or \designSpaceItem{Application}-level option (\figref{fig:context-models-sidepanel}'s Side-panel).
These designs form a spectrum of implicit (does not inform the user of the model's context) to explicit (allows the user to specify the code context) context designs.

\paragraph{Disambiguation} As described above, the ambiguities latent to specification incompleteness and model uncertainty can prompt some interfaces to  produce multiple candidate solutions where the end-user needs to select a correct solution.
Prior studies have indicated that disambiguation can be valuable~\cite{barke2022grounded, mayer2015user, li2020multi}, but there has been little consideration of the types of UI affordances that might support such work. To this end, we considered: \emph{What types of disambiguation representation do users perceive as valuable?}
We presented a sequence of design options guided by prior explorations, shown in \figref{fig:disambig}. These included
(A) an Output Only display inspired by \surveySystem{VizSmith}~\cite{bavishi2021vizsmith} (\designSpaceItem{Effects}),
(B) a Code Only display inspired by \surveySystem{Copilot}'s multiple suggestions in a new tab feature~\cite{copilotAlts} (\designSpaceItem{Code}),
(C) a Code with Diff display per \citet{barke2022grounded} (\designSpaceItem{Code} and (\designSpaceItem{Summary}),
(D) a  Code and Output display (\designSpaceItem{Code} and \designSpaceItem{Effects}),
(E) a higher-level Summary View (\designSpaceItem{Summary}), and a paginated view (\designSpaceItem{No Disambiguation}) as in \figref{fig:context-models-inline}.
\paragraph{Adaptation} While contemporary AIs are powerful, their results are often imperfect~\cite{weisz2021perfection} due to model limitations or ambiguous user specifications.
Users often accept these imperfect solutions and then engage in an accept-validate-repair sequence~\cite{barke2022grounded} as a way to read, internalize, and adapt the generated code to the local interface. Here we examined: \emph{How do users expect the interface to assist in the adaptation of generated code?}
We highlight that \emph{adaptation} is a different process than \emph{refinement}, involving the use of generated code as opposed to getting the assistant to produce better code.
We show design options that sought to elicit the adaptation strategies,
as in \figref{fig:fixing-code}, which variously surface \designSpaceItem{Model Metadata} and Disambiguation \designSpaceItem{Summaries}.
These include (A) a linter-style UI that highlights potential semantic-level mistakes in generated code,
(B) an option where the assistant generates a Snippet skeleton (instead of executable code) where the user is prompted to fill in missing items (per \citet{barke2022grounded}),
(C) a Token-confidence design that shows model confidence on different parts of the code,
and (D) a Token-alternatives option that lets the user explore alternatives for a given part of the code.
(C) and (D) are from \citet{weisz2021perfection}.

\paragraph{Provenance}
A touted value of computational notebooks is their facility to interleave documentation and code in a literate manner~\cite{rule2018exploration, kery2018story}, which allows for description of code and design variations and explorations~\cite{wood2018design}.
\citet{barke2022grounded} suggest that interactions with \surveySystem{Copilot}-style interfaces are expected to be transitory, which is aligned with nearly every interface examined in \secref{sec:design-space}---\surveySystem{B2} excepted.
Disentangling these seemingly conflicting positions is an important step toward understanding how to develop tools in this environment.
Here, we considered \emph{What, if any, documentation of code \designSpaceItem{Provenance} of generated code do users believe is valuable?}
Our designs examining this tension included an ``autosummary'' button that saves the interactions with a code assistant as either a markdown cell or as an inline comment, a cell-level marker to designate which cells contain generated code, and an explorable log that documents the interactions with assistants (per~\citet{kery2017variolite}).
Figures showing these designs are in the appendix.

\section{Interview Study Analysis}
\label{sec:analysis}

We now reflect on the themes, topics, and suggestions that we identified in the interviews. We begin with a summary and then present cross-cutting themes (and return to the particulars of our UI results), ordered from most specific to most general.
These include
the role the UI design has on the perception and usage of features in this context (\secref{sec:ui-design}),
the relationship that a code assistant in a notebook has with its surroundings (\secref{sec:surroundings}),
and finally, the role that code suggestions have on trust (\secref{sec:trust}).

All participants believed it is valuable to adapt code assistants to notebooks.
For instance, \pX{11} noted \pQalt{the use case for [code assistants] in notebooks is, I would say, pretty strong}.
\pX{4} saw
\pQalt{a lot of potential where you can go  with [code assistants] for the data scientists.}

Yet, for most topics we considered, there was no consensus about a single best feature over other design choices for code assistants.
No one of the interfaces we described for code context was universally preferred,  with each being liked or disliked according to participants' preferences for clutter, politeness, and locality.
Participants nearly unanimously viewed Code and Output (\figref{fig:disambig}D) as the best way to explore collections of suggestions, although some participants (\eg{} \pX{4} and \pX{11}) liked the Summary View---likely for its novelty.
Each of the presented refinement strategies---excluding Token-confidence---was seen as valuable for different reasons ranging from familiarity (\pX{13}) to transparency (\pX{4}).
Artifacts of interactions with code assistants were seen as unnecessary, as code generation was not viewed as a part of their deliverable.
In concert, this suggests a wealth of opportunities in this design space to serve differing desires and interests, as well as the wickedness~\cite{rittel1974wicked} of the problem---given the multi-faceted and contradicting ways in which data scientists expect to use code assistants and notebooks.
As notebooks are a subset of interfaces more generally, many of our findings align with similar expectations users might have of other contexts. However, our results are specific only to the context of code assistants in notebooks as understood by data scientists, as we only sought opinions from this type of user in this context.
We reference these topics as code assistants and users, respectively, as a notational convenience, but they should be understood as these specific cases rather than general ones.

\subsection{Role of UI Design}
\label{sec:ui-design}

The way that users can interact with code assistants shapes what they attempt to do with those assistants. Here we consider how the design of the interface itself might influence different behaviors.

\begin{figure*}[t]
    \centering
    \includegraphics[width=0.8\linewidth]{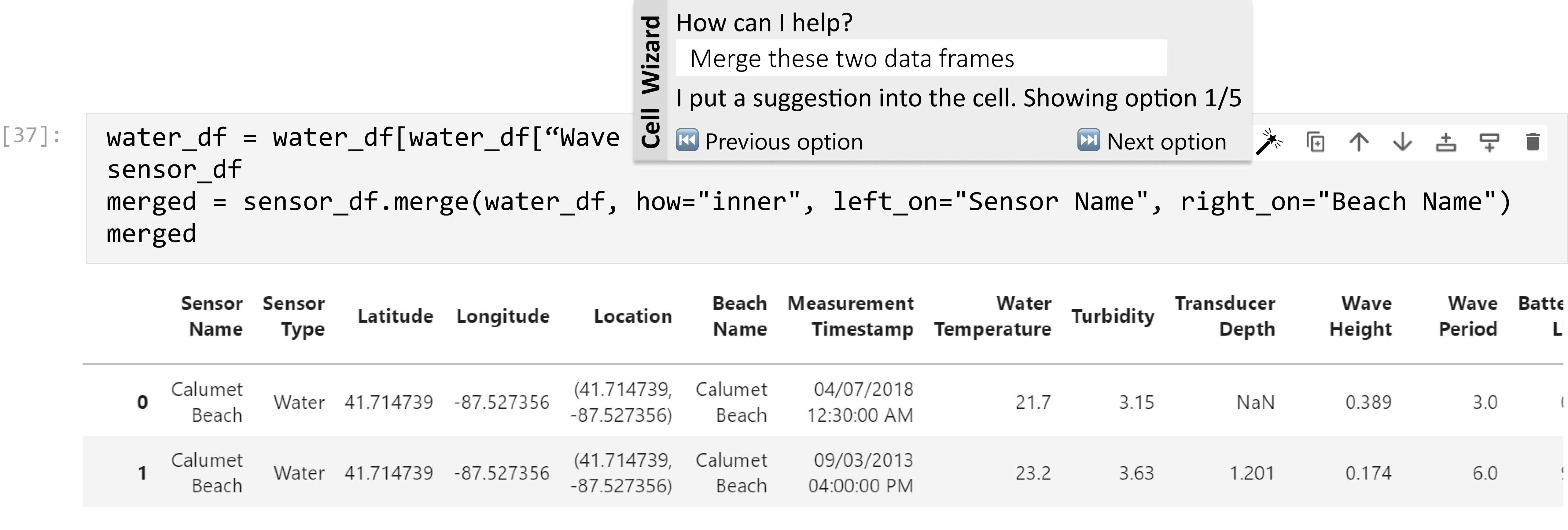}
    \vspace{-1em}
    \caption{The Wizard interface (\designSpaceItem{Targeted cell}) included in our interviews surfaces control of a code assistant via a popover specific to each cell. The context used to inform the code generation is specified implicitly and is local to the targeted cell.
    }
    \label{fig:context-models-wizard}
    \Description{A screenshot of the proposed wizard design, which is a normal cell with a popover in which the user can input requests.}
\end{figure*}
\begin{figure*}[t]
    \centering
    \includegraphics[width=0.9\linewidth]{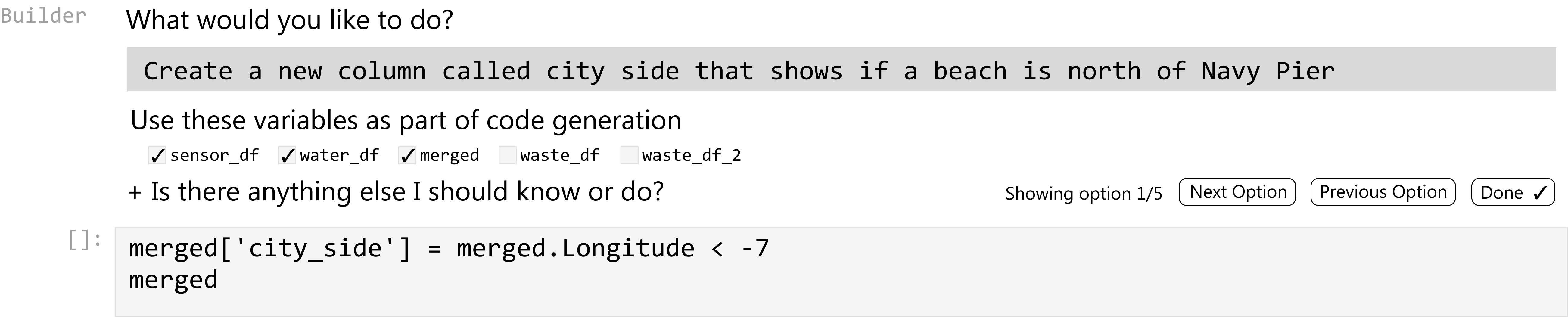}
    \caption{The Inline interface (\designSpaceItem{Inline}) places an assistant between cells, akin to any other notebook cell. Code generation context is formed by selecting variables that should be used to form the scope, however other strategies might also be used.
    }
    \Description{A screenshot of the proposed inline design, in which a new type of notebook cell has been added beneath the familiar ones in which the user can type a natural language request.}
    \label{fig:context-models-inline}
\end{figure*}

\subsubsection{Context and Interfaces}
\label{sec:context-interfaces}
In presenting several different modalities for interacting with code assistants, we sought to elicit perceptions about what and how much context participants believed would be useful for code generation tasks.

\begin{figure*}[t]
    \centering
    \includegraphics[width=0.9\linewidth]{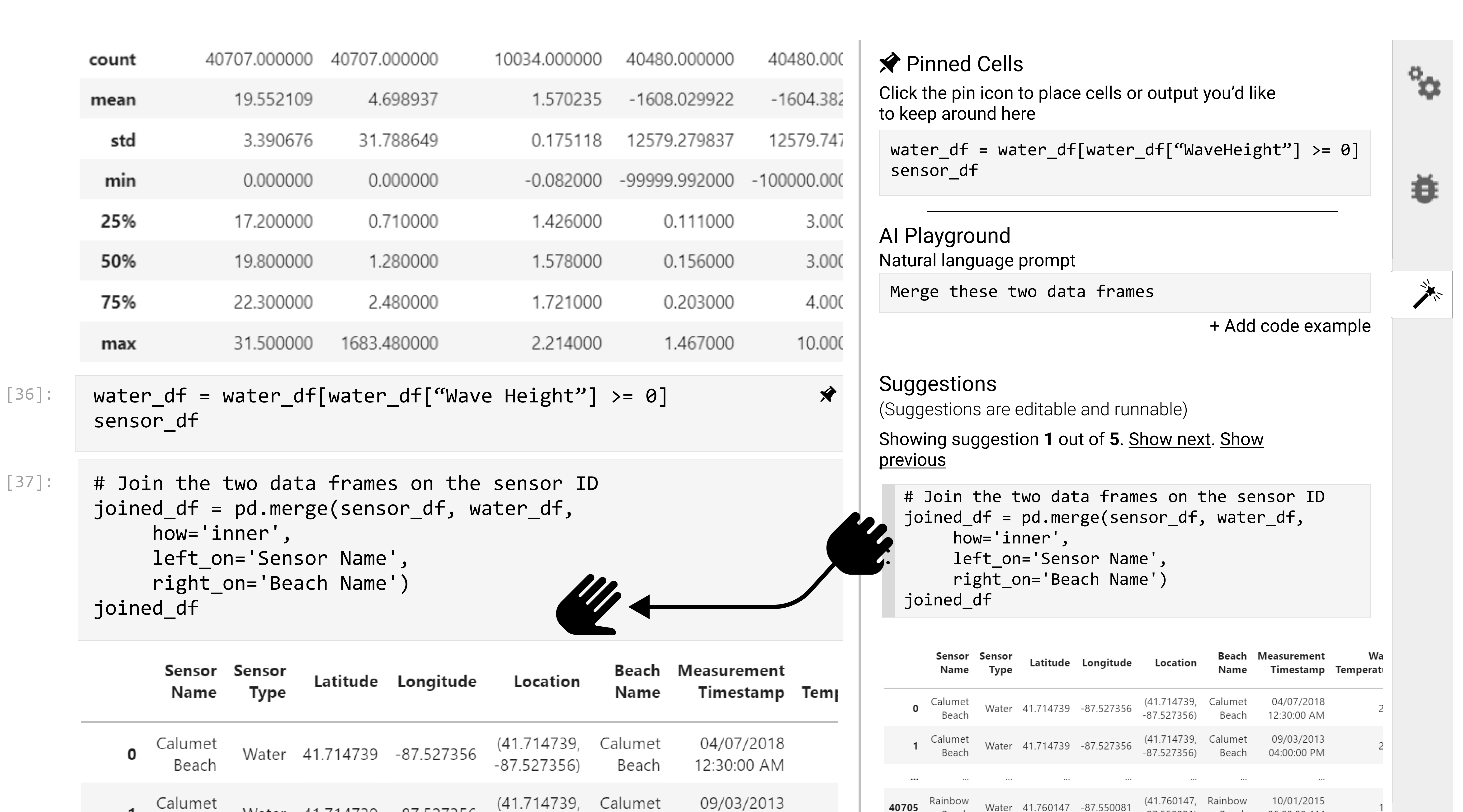}
    \vspace{-1em}
    \caption{
        The Side-panel interface (\designSpaceItem{Application}) paradigm sits outside of the notebook flow. It provides a sandbox to explore suggestions before dragging them in. Context is described explicitly by pinning cells or through code or natural language.
    }
    \label{fig:context-models-sidepanel}
    \Description{A screenshot of the proposed side panel design, which is a panel parallel to the notebook in which the user can make requests to the model via natural language and pin notebook cells into it to create a context for code generation.    }
\end{figure*}

Most participants did not care about manual specification of context,
typically expressing \pQ{8}{I want [the code assistant] to look at everything,} rather than requiring manual inclusion or exclusion of parts of the code.
Some participants valued having control over what goes into code generation (\eg{} \pX{1}, \pX{6}, and \pX{7}), but those participants also had non-trivial knowledge of the underlying code-generating model.
For instance, \pX{1} \pQalt{liked the idea of selecting variables for the code generation.}
This group suggested explicit opt-in inclusion of particular cells (\pX{1} and \pX{6}) and giving priorities to certain gestures---such as library usage (\pX{7})---might be valuable.
\pX{6} and \pX{13} said context-free suggestions (akin to a search functionality) would be useful---echoing \citet{barke2022grounded}.
Some participants liked the \designSpaceItem{Inline Cell} design, as in \figref{fig:context-models-inline}, because it was familiar (\pX{1}, \pX{4}, \pX{7}, and \pX{15}) and immediate (\pX{2}) and believed it would allow them to keep a consistent mental model of the execution flow.
However, \pX{10} and \pX{13} disliked Inline because they believed that it cluttered the notebook and would make it hard to keep track of interactions with the assistant.
Others preferred the Side-panel (\designSpaceItem{Application}), as in \figref{fig:context-models-sidepanel}, because of the ability to keep things separated. \pX{5} noted that \pQalt{I don't like to mess with my code...unless I'm sure} about the changes to it.
\pX{4}, \pX{5}, and \pX{13} noted that they tended to have large screens and so having a separate section of the notebook dedicated to interactions with code assistants was a good use of that space.
\pX{4} analogized it to using an integrated Google-like search functionality.
However, some (\eg{} \pX{2}, \pX{8}, and \pX{15}) disliked the perceived ergonomics of needing to split their attention between areas of the notebook.
\pX{14} and \pX{15} pointed to the Wizard design (\designSpaceItem{Targeted Cell}), as in \figref{fig:context-models-wizard}, as being valuable because of its locality.
Others were averse to the Wizard because of a dislike of popups:
\pX{3}, \pX{4}, \pX{7}, and \pX{12} felt that it would \pQ{4}{break my train of thoughts every time} or bring them out of \pQ{7}{the zone}.
Only \pX{10}, \pX{11}, and \pX{14} brought up the autocomplete style found in \surveySystem{Copilot} as being a desirable alternative to the designs we presented---although \pX{11} specifically noted that he would prefer them to be used in conjunction, as they served different purposes.
For instance, \pX{11} and \pX{15} seemed to think of creating visualizations as a different process than other forms of coding, \pX{15} noting that \pQalt{it doesn't add value to your final product, like data wrangling [does].}
This suggests that there are numerous design opportunities to address tasks that users do not view as central to their contribution to the work.

\begin{figure}[t]
    \centering
    \includegraphics[width=\linewidth]{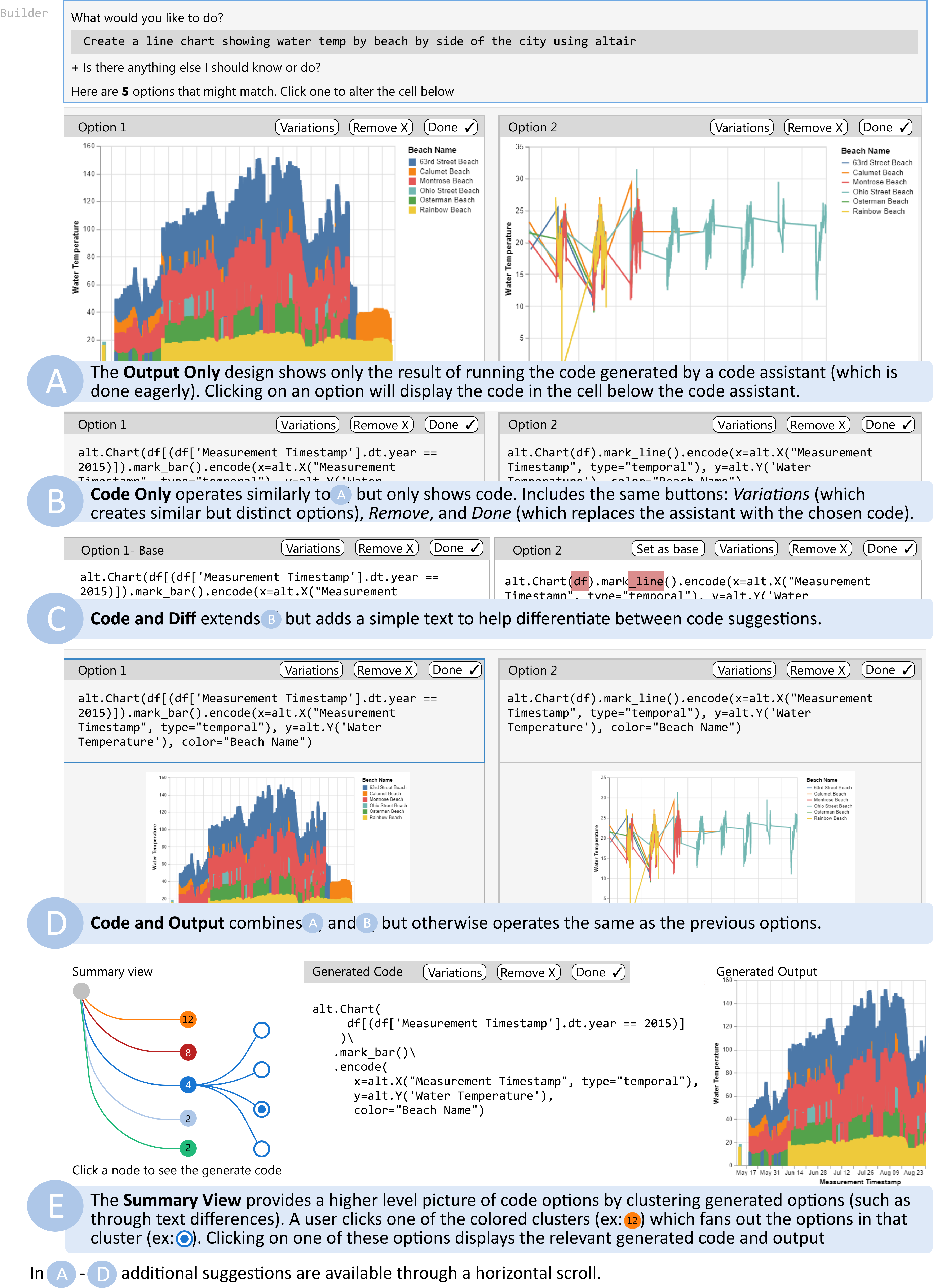}

    \caption{We presented a variety of options for disambiguation covering combinations of \designSpaceItem{Code}, \designSpaceItem{Effect}, and \designSpaceItem{Summary} strategies.
        These options could be applied to other styles besides \designSpaceItem{Inline} (as here), such as \figref{fig:context-models-wizard} or \figref{fig:context-models-sidepanel}.
    }
    \Description{A set of 5 screenshots arrayed vertically showing the different designs. (A) shows a row of generated charts beneath a cell labeled "output only". (B) shows a row of generated code snippets beneath a cell and is labeled "code only". (C) shows the same content as (B), but a color-based annotation has been added to highlight their textual differences. (D) shows the combination of (A) and (B). (E) shows a graph with colored edges labeled "Summary view" that also shows a single code snippet and chart.}
    \label{fig:disambig}
\end{figure}

Many participants (\pX{2}, \pX{4}, \pX{6}, \pX{8}, \pX{11}, \pX{13}, and \pX{15}) noted that their usage of a code assistant was not motivated by a desire to learn or explore (although \pX{11} and \pX{14} espoused this view), but simply increase their efficiency.
This focus on efficiency is in tension with some tasks at the heart of notebooks' literate programming style. One such example is data analysis, which may require alternating modalities to investigate correctness, such as interacting with a chart or spreadsheet to determine if a data manipulation has been executed correctly.
Features intended to support exploration may be able to take greater liberties in their designs (which \pX{3}, \pX{7}, and \pX{11} noted about \figref{fig:disambig}E's summary view), while those focused on acceleration require a shorter and less-noticeable interaction loop.

Evidently, not all designs will work for all users. It may be beneficial to surface multiple places in the interface where the user can interact with code assistants---akin to how spelling/grammar checkers are often surfaced through both a triggerless~\cite{Jayagopal22Exploring} ambient mode as well as a mode triggered from a menu.

\subsubsection{Feature Prominence}
\label{sec:prominence}
A paradox of interface design is how to make something discoverable---and easy to remember---without making it annoying.
Demo or novelty-driven design can be helpful for buy-in but can make interactions with those systems frustrating in practice---as was the infamous case of Microsoft Clippy~\cite{Correll21Clippy}.

\begin{figure}[t]
    \centering
    \includegraphics[width=\linewidth]{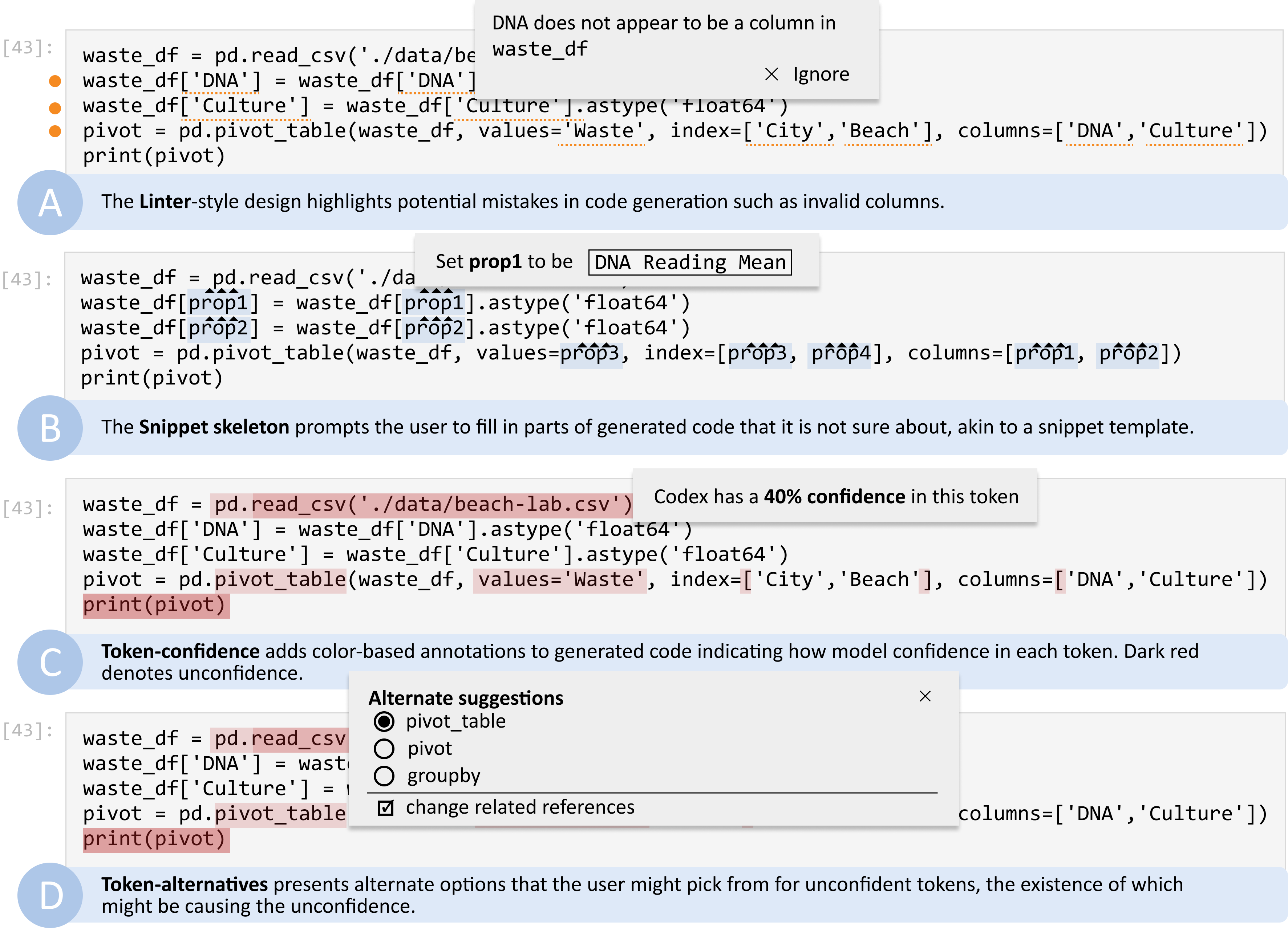}
    \caption{The four designs presented to participants in our study for adapting code generated by an assistant to the task or domain.
    }
    \label{fig:fixing-code}
    \Description{One screenshot of a notebook cell repeated four times showing proposed features. (A) adds what appears to be a spell checker and is labeled "linter". (B) add a fill-in-the-blank system for a series of values throughout the snippet and is labeled "snippet skeleton". (C) adds colored patches that are gradients of red over the code and are labeled "Token-confidence". (D) adds a clickable menu to (C) that allows the user to select from several alternatives for that section and is labeled "Token-alternatives".}
\end{figure}

A recurring issue in the interviews related to how prominent the controls for a coding assistant were. A feature that is too prominent might be \pQ{15}{irritating}, while one that is too subtle will be \pQ{9}{undiscoverable} or unlearnable.
A feature being prominent means that it is more likely to be discovered but also potentially that it will be disabled (\pX{15}).
\pX{3} valued prominence, noting it prompted him \pQalt{to pay attention to this new feature,}
whereas features that do not announce themselves may not get found. For instance, only \pX{1}, \pX{7}, and \pX{10} were aware of \surveySystem{Copilot}'s alternatives menu~\cite{copilotAlts}.
\pX{9} noted that he preferred the Inline design because it was easier to discover compared to the Wizard or Side-panel (an advantage shared by triggerless initiations~\cite{Jayagopal22Exploring}), whereas \pX{4} noted that he preferred the Side-panel because it could more easily be ignored when it was not relevant.
Yet, triggerless systems are not appropriate solutions for all cases: \pX{11} commented about \surveySystem{Copilot} that it was \pQalt{frustrating for me to kind of say, like, No, I don't want this right, because it kept on trying to auto guess}---a position also shared by \pX{13}.
UIs that abuse triggerless or eager presentation may poison the well for users. For instance, \pX{15} noted that every time \pQalt{I see an attempt from an interface to help me, I just disable it,} having had negative previous experiences.
\pX{7} and \pX{15} suggested that interfaces that may sometimes be annoying (referring to \figref{fig:disambig}C's diff view and \figref{fig:fixing-code}C's token-confidence view)  might be more usefully shown on demand rather than all of the time, that is, as lenses---echoing Kery \etals{}~\cite{kery2018story} recommendations for debugging or history tasks in notebooks.
This style of lightweight opt-in feature may be useful,
however, this may make that feature all but invisible.
We suggest then that navigating the tension between being polite~\cite{whitworth2005polite} (by respecting user agency) and promoting usage (such as finding opportunities for usage that could be missed otherwise) is central to the design of effective code assistants.
Familiarity with and novelty of particular features were powerful factors in participants' expected usage patterns.
\pX{9} discovered the snippet search functionality in his editor by using a search engine to investigate if \pQalt{there is anything like [snippet search] by PyCharm?}
\citet{Jayagopal22Exploring} note that this is a factor for novices, but the presence of such concerns among our participants suggests it may be more universal among notebook users.
For instance, \pX{6} noted that his use of Token-confidence displays (\figref{fig:fixing-code}C) in a past project biased him toward them.
\pX{5} liked the output-only (\figref{fig:disambig}A) displays because they reminded her of related features in Excel.
Such expectations are closely informed by the vocabulary of features found in other notebooks (\cf{} \secref{sec:domain}).
The summary view feature (\figref{fig:disambig}E) was sometimes viewed as intriguing (\eg{} \pX{2} and \pX{3}), however, that may be due to novelty.
Novelty and breaking expectations in surprising ways can be beneficial and lead to magical feeling experiences---although, as prior studies on \surveySystem{Copilot} have shown, the magnitude of this improvement may be overestimated by end-users~\cite{imai2022github, vaithilingam2022expectation}.
Participants observed that novelty can have value (in that it can provide an incentive to explore or interact with a feature), but the \pQ{11}{value of [the feature] has to outweigh the novelty}.
We suggest that, as with humor~\cite{warren2021makes},  \lesson{expectation can be subverted to useful and surprising effect, however, such manipulations can be seen as annoying and lead to feature disuse.}

\subsection{Relationship with surroundings}
\label{sec:surroundings}

Notebooks do not exist as solitary objects. They serve a broad range of purposes, including exploration and experimentation (\pX{4}, \pX{7}, and \pX{9}), an environment for development before formalization into a script or pipeline (\pX{4}), presentation or report (\pX{3} and \pX{13}),
as well as both shared (\pX{5} and \pX{7}) and solitary objects (\pX{2}, \pX{6}, and \pX{8}).
We found that being situated in such usages informs the desired UI affordances, such as for contextualization in general and domain-specific enhancements particularly.

\subsubsection{Provenance}
\label{sec:documentation}
Participants did not view documentation of interaction with code assistants (\designSpaceItem{Provenance}) as important.
For instance, \pX{1}, \pX{2}, \pX{4}, \pX{5}, \pX{8}, \pX{9}, \pX{10}, \pX{11}, \pX{12}, and \pX{15} noted that they only cared about the quality of the output and not how it was achieved, with some participants (\pX{3}, \pX{5}, \pX{6}, \pX{8}, \pX{9}, \pX{11}, and \pX{13}) adding that data science is an output-driven discipline.
\pX{13} noted that data scientists \pQalt{spend more time thinking about insights, thinking about strategies, rather than thinking about how to code,} highlighting that tasks like documenting code provenance are not seen as essential to their role.

Some participants viewed our documentation designs as only useful in untrusted situations (\pX{2}, \pX{5}, \pX{8}, and \pX{13}), which were uncommon in their work. For instance, \pX{2}, \pX{5}, \pX{7}, \pX{11}, and \pX{13} noted that they trusted any notebooks presented to them by a colleague and thus would not need to audit the code's origin.

\pX{1}, \pX{4}, \pX{5}, and \pX{13} compared documentation of code generation as being similar to including attribution to Stack Overflow posts from which they may have copy-pasted code. \pX{1} observed that \pQalt{Stack Overflow is like a really slow coding assistant}.
Most participants did not attribute code gathered from other sources because they did not view it as something worth documenting---unless it was particularly unusual code or they might not have been expected to know it (\pX{14}).
\pX{8} and \pX{15} expressed skepticism for the utility of any attribution at all, with \pX{8} questioning \pQalt{What is human's memory and creativity? How did you know whatever you write is even your original thing or if it exists somewhere else?}
These reservations agree with prior observations~\cite{epperson2022strategies} that data scientists tend not to see code as a deliverable.

While hesitancy about documentation may be surprising---in light of the well-known~\cite{kery2018story, wood2018design, rule2018exploration} and well-discussed (\pX{1}, \pX{2}, \pX{4}, \pX{7}, and \pX{8}) literate programming paradigm in which notebooks operate---
this disinterest seems to suggest that there are different categories of work in notebooks: that which is meaningful to the deliverable and that which is only necessary for its construction.
For instance, \pX{9} suggested that documentation of interactions with code assistants would be something akin to recording or publishing the undo/redo stacks associated with typing.
Prior work~\cite{barke2022grounded,Sarkar22WhatIsItLike} suggests that documentation of interactions with code assistants is not seen as valuable, and we extend this to suggest that \lesson{it may be beneficial to  specifically align the design of code assistants so that they fall into the category of work that is not viewed as significant to document} because, by being ignorable, they can more easily become commonplace---as is the case with spell checkers. That is, they should be treated like a ghostwriter.

\subsubsection{Search}
\label{sec:search}
It has been argued that integration of search with code suggestions improves the facilities of code assistants.
For instance, \citet{xu2022ide} found that code search enabled different categories of tasks, such that users employed search for larger or more complex pieces of code and synthesis for small simple modifications.
Similarly, prior work has explored integrating search into IDEs~\cite{brandt2010example}, as well as more specifically in notebooks~\cite{li2021edassistant, li2021nbsearch}.

Our participants (\eg{} \pX{4}, \pX{6}, \pX{8}, \pX{13}, and \pX{14}) espoused a similar set of desires, describing that code search may be a valuable addition to code assistants.
This is in line with how some users already expect to use code assistants~\cite{Sarkar22WhatIsItLike}, for instance, \pX{4} noted that he already tended to think of \surveySystem{Copilot} as a search system.
\pX{9} spoke at length about the value of being able to search against a fixed library of snippets---although this library also might be usefully made malleable.
For instance, \pX{13} wanted to be able to save snippets (such as her preferred manner of cleaning categorical data)
and have them be suggested later.
\pX{8} and \pX{14} wanted to be able to access elements that were outside of the model's training data, such as code written by their teammates or others within their organization, but have it still be adapted to context rather than shown in a decontextualized way.
As \pX{3} noted that \pQalt{if I know my code is going to be shared across the team, I usually tend to write a lot of comments} (echoing prior work~\cite{epperson2022strategies}), code search with automated integration may reduce the need for manual documentation.

Yet search alone was not sufficient: it seems that it is necessary to also provide context or reasoning for code suggestions.
For instance, \pX{5} desired that context should be integrated into each suggestion, such as through library documentation.
\pX{9} suggested a similar desire for \pQalt{documentation for each [of the suggestions], then you can look into it like, what are the pros and cons of it.}
\pX{2} and \pX{5} desired a way to see what parts of the training data influenced each code suggestion to get a better understanding of how that code was used---a feature commercialized in CodeWhisperer~\cite{amazonCodeWhiperer}.
Prior work~\cite{rong2016codemend, vaithilingam2022expectation} also has recommended library documentation be included as part of code recommendation, which is an instance of the Human-AI guideline that systems should ``make clear why the system did what it did''~\cite{amershi2019guidelines}.

We suggest that \lesson{integration of both search {\emph{and}} documentation into code assistants is valuable} as it can help users find code and validate it.
We note that these contextualizations (or similar \designSpaceItem{Model Metadata}) may increase automation biases~\cite{vaithilingam2022expectation} and move users to be even less questioning~\cite{mcnutt2020divining} of suggestions.

\subsubsection{Relationship to domain}
\label{sec:domain}
Critical to any usage is the context or domain in which that work is done. We identify several ways in which knowledge of the domain might aid tool design.
We found that participants' expectations were guided by the makeup of the data science ecosystem.
For instance, analysts often~\cite{epperson2022strategies} conduct analyses in environments in which they do not have full control, like Databricks, in order to use protected resources, like Spark clusters.
Such systems often provide their own variant of notebooks with their own collection of UI affordances drawn from a relatively limited vocabulary of allowed interface forms in notebooks~\cite{lau2020design}.
While \pX{14} noted it can be easy enough to port from one system to another, \pX{13} commented that she uninstalled \surveySystem{Copilot} after months of usage noting that
\pQalt{I don't want to be too reliant on it,}
as it was not available in all of the environments in which she needed to do work.
A motivated user might adopt new interaction forms specific to a particular environment, but, as \pX{15} argued, users whose key concern is efficiency may quickly dismiss things they perceive as hampering their process.
Thus \lesson{code assistants may be more likely to be adopted if they follow or are in dialogue with familiar patterns}.
We suggest that it may be useful to build code assistants that operate on a browser level rather than a notebook level to take advantage of the necessity of moving between browser-based notebook environments.

As code assistants generate code, it is natural to consider how they relate to best practices and code styles.
Some participants (\pX{6}, \pX{10}, \pX{13}, and \pX{15}) thought that it would be better if assistants adapted to their style and learned their preferred way of doing things over time.
In contrast, \pX{14} noted that he would not want a system to adapt to his style because \pQalt{I don't want to stay in the bubble.}
Similarly, \pX{12} noted a preference to exert control over \pQalt{naming convention and those kinds of stuff.}
Others (\pX{5}, \pX{7}, \pX{8}, \pX{9}, and \pX{11}) thought that generated code should strive to follow best practices and  match the conventions of their team.
\pX{4} believed that code assistants helped enforce and teach best practices more effectively than traditional assistants such as linters or auto-formatters, observing \pQalt{for me, Copilot is just an evolution of these tools.}
This is related to Sarkar \etals{} \cite{Sarkar22WhatIsItLike} observation that some users believe that \surveySystem{Copilot} helps with best practices.
\lesson{It may be useful to capitalize on this expectation of offering best practices as a way to integrate opinionated advice, although this should be done cautiously as automation bias~\cite{vaithilingam2022expectation} may reinforce negative behaviors}.
A feature of frequent interest (\pX{1}, \pX{3}, \pX{5}, \pX{7}, \pX{9}, and \pX{10}) was a \emph{linter for notebooks}.
Such a system could assist with common data science errors such as out-of-order execution (\pX{1}), side effects (\pX{10}), and identify opportunities to convert frequently rerun cells to functions (\pX{5}) in a manner analogous to how spell checkers offer suggestions on grammar or usage errors.
Notably, this typically came up prior to discussion of \figref{fig:fixing-code}A's linter-style design for adaptation.
A common observation was that code in notebooks tends to be lower quality---\pQ{5}{I guess I'm lazier when I'm using a notebook}---suggesting that an ambient design intervention, like a linter, agrees with typical usage patterns in notebooks.
However, this UI style can be seen as annoying if it does not provide consistent utility: \pQ{7}{sometimes linter just complains, it just says, too many lines or too many variables within that function}.
In addition to providing basic notebook usage hints (such as those highlighted by \citet{rule2019ten}), it may be useful to support domain-specific practices such as schema awareness (\pX{1}, \pX{5}, \pX{12}, and \pX{14})---like highlighting when a column does not exist on a dataframe or when a particular method is slow compared to a vectorized alternative---or best practices local to data science---like ensuring that data fitting uses a train-test split pattern.
Domain-specific linters have been created in other domains---including visualization~\cite{wu2022ai, chen2021vizlinter, mcnutt_surfacing_2020}, spreadsheets~\cite{barowy2018excelint}, and ML data~\cite{hynes2017data}---\lesson{suggesting the value of lightweight domain or medium-specific ambient assistants for data science in notebooks}.
While not addressing every domain concern, such a lightweight assistant may reduce the need to specifically adapt more-powerful coding assistants to notebooks.

\subsection{Role of trust and control}
\label{sec:trust}

We found that the relationships that participants have or expect to have with code assistants are closely mediated
by their sense of control---how they understand what it does and their ability to accept or reject suggestions.
These aspects are mediated by perceived quality of output (including correctness and readability) and knowledge (or perception) of the underlying model.

\subsubsection{Points of control}
\label{sec:control}
The primary point where users can exert agency in our \emph{specification-refinement} loop is through gesture and interpretation of output.
\pX{1} noted that \pQalt{being in control of the code} felt essential to trust in the tool.

We found that a critical point of control is the way that participants expected the assistant to integrate into their workflow.
\citet{barke2022grounded} observed acceleration and exploration modes in \surveySystem{Copilot} usage in which users either used it as a way to simply type more quickly or to try to more generally understand how to do something.
Participants (\eg{} \pX{1}, \pX{2}, \pX{7}, \pX{8}, \pX{10}, \pX{14}) believed they only used  assistants like \surveySystem{Copilot} to increase their efficiency (akin to advanced autocomplete).
Participants did not directly self-describe that they used \surveySystem{Copilot} to explore alternatives, for instance, \pX{8} noted that he did not \pQalt{because I don't really trust it} to know uncommon functions or libraries.
\pX{5} observed that type or syntax-based autocomplete engines found in many editors are sufficient for exploration of libraries and function parameter values.
Yet some participants did seem to value the results of using code assistants for exploration.
For instance, \pX{2}, \pX{4}, and \pX{9} described it as a way to help them learn to do new things, while \pX{6} and \pX{14} observed that it helped them learn new coding patterns.
Our findings thus comport with \citet{barke2022grounded} regarding bimodal usage, however, our view of this behavior is more limited as we only have participant self-reports rather than direct observations.

Some participants wished to be able to control certain parameters (\designSpaceItem{Model Metadata}) about the assistant, such as \pQ{8}{suggestion length} or \pQ{1}{adventurousness}
(\ie{} more exploratory or less straightforward suggestions, akin to increasing the model temperature).
\pX{12} expressed an interest in being able to swap between models (\designSpaceItem{Model-Agnostic}) if he could see an automated report of his use of each model, while \pX{11} desired a toggle to indicate whether or not the system should try to help learn rather than merely complete the task.
While this sort of configuration can be valuable, many participants noted that it (and other presentations of choices) can be overwhelming (\pX{2}, \pX{7}, \pX{8}, and \pX{13}).
Hiding these choices in an expert menu may reduce this burden, but it also may cause those elements to remain undiscovered (\secref{sec:prominence}), even by those who might wish to use them.

Despite the seemingly static nature of code, its role as a component of an editor is inherently dynamic and, in this context, involves both reading and adaptation as points of control.
For instance, \pX{1} noted that he felt out of his control
when he could not understand generated code (\eg{} when \surveySystem{Copilot} suggests hard-to-understand one-liners).
\pX{3} expressed some hesitancy about generated code more generally: \pQalt{I'm not like somebody who can only read code. I do write code and I do want to have the control.}
\citet{drosos2020wrex} and \citet{kery2020mage} highlight the importance of synthesizing readable code in contexts like data wrangling, while \citet{Sarkar22WhatIsItLike} make similar observations specifically related to \surveySystem{Copilot}.
Following \citet{weisz2021perfection} some participants thought that code annotations would help them better read and understand suggestions.
For instance, \pX{5} noted that \pQalt{reading someone else's code is always harder} and that design interventions such as those presented in the study would \pQalt{definitely be helpful to read [generated] code.}
However, \pX{3} and \pX{9} thought annotations, such as diffs or Weisz \etals{} \cite{weisz2021perfection} designs, would be distracting.
Participants described a variety of adaptation strategies, including rewriting the prompt (or just continuing to type in the case of triggerless systems like \surveySystem{Copilot}), fixing the output (following Barke \etals{}~\cite{barke2022grounded} accept-validate-repair sequence), or simply completing the task manually---although the use of these strategies was mediated by experience with the language (\pX{5}) and estimated time to fix the errors (\pX{4} and \pX{12}).
Our interviews explored design interventions that might augment these strategies, although participants were divided on which particular flavor would be valuable.
For instance, \pX{1}, \pX{4}, and \pX{10} liked the Snippet skeleton approach because it felt like they were able to exert agency over boilerplate,  with \pX{4} noting that \pQalt{For me, the code skeleton is the best one because you are being transparent and honest.}
\pX{5} and \pX{13} liked the linter-style approach because it was familiar from other programming contexts.
\pX{9} found numeric representations of confidence to be confusing (as in \figref{fig:fixing-code}C's token-confidence view), noting that \pQalt{that's not useful to me. For me, it's either 100\% or zero.}
This potentially disagrees with prior work, which found that numeric heuristics as being associated with algorithmic intelligence~\cite{Liao22Designing}---although this may be due to our participant population, data scientists, who may have higher computational literacy than other groups.
We suggest that \lesson{effective code annotation in this context should either be ambient (ignorable unless needed) or provide value for active interaction (as a dedicated input mechanism)}.

\subsubsection{The effects of knowledge}
\label{sec:knowledge}

We observed that the process of understanding the capabilities of a code assistant is one of forming a relationship.
Participants with prior relationships with such models bring with them expectations that in turn inform their usage.

Participants with experience with LLMs claimed that they would be tolerant of bad answers.
\pX{8} and \pX{14} noted \surveySystem{Copilot} was only useful for problems that were well represented in the training data. \pX{8} highlighted them as being limited to popular APIs (such as NumPy or pandas) and that more esoteric topics or APIs that change rapidly tend to induce incorrect results when using \surveySystem{Copilot}---a point also noted by \citet{Sarkar22WhatIsItLike}.
Similarly, \pX{12} believed that \pQalt{I can tolerate its wrong answer. But I may generalize the type of mistake that it makes, like, for example, if it's bad at generating graphs.}
Participants with a less clear understanding of the systems underlying \surveySystem{Copilot} believed their reactions to poor performance would be more severe, noting that after two or three bad suggestions that would probably turn it off (\pX{3}, \pX{9}, \pX{11}, and \pX{15}) or \pQ{5}{if it's not too overbearing, I might just ignore it}---highlighting the potential advantages of polite triggerless designs.
It seems that these first impressions are an important part of developing the working relationship with the assistant: if value or trust is not clearly established either through preexisting assumptions or early behavior, then users may be unlikely to continue to try to use it.

\citet{yildirim2022experienced} note that if designers have at least a limited understanding of what AI can do, they are better able to use it as a design material. \citet{lau2008pbd} highlights a similar guideline in the design of effective synthesis tools. \citet{ferdowsifard2020small} observe the ``user-synth gap'' in which it is not clear what the synthesizer can produce, which can cause an impedance mismatch between user behavior and expectation, which is a case of the Human-AI guideline~\cite{amershi2019guidelines} to clearly communicate system abilities.
We suggest that similar principles may apply to users of notebook-based code assistants, as \lesson{giving users a model of what is going on under the hood will likely shape their approaches to understanding and using generated code and affordances therein}.

\subsubsection{Verification}
\label{sec:verification}
Even if a provided solution looks right, it may need to be examined to ensure its correctness.
Participants described how the need for verification is tempered by how high stakes the code is, how long it will take to run, and the perception of its quality.

In addition to code reading, participants also described employing strategies like consulting documentation as a way to verify suggestion correctness.
For instance, \pX{6} noted that \pQalt{I still Google to verify its correctness} in cases when it would take multiple minutes to establish whether or not a system worked correctly.
\pX{8} noted \pQalt{I mean, I always double-check} when using an uncommon API, such as for creating ML pipelines.
\pX{2} and \pX{14} desired validity checks akin to popularity markers in other settings.
Some noted that having a low friction way to explore suggestions (as in \figref{fig:context-models-sidepanel}'s sandbox) before accepting them would enhance their trust in the model:
\pQ{8}{I think it will make me feel more comfortable with...trying out longer suggestions and accepting it}.
\pX{1} added that such features \pQalt{make me feel safer.}
Others (\pX{7}, \pX{9}, and \pX{10}) felt that this was unnecessary and could be achieved through normal cell usage.

A common means of verification explored in our interviews was through the exploration of alternative suggestions (\ie{} \designSpaceItem{Code} or \designSpaceItem{Effects} disambiguation).
This allowed them to explore variations on different approaches to different tasks, such as visualization or particular activities like identifying \pQ{9}{different ways to connect to the blob,} and in doing so verify the output of the assistant.
Valuing browsing agrees with prior observations of \surveySystem{Copilot} users~\cite{barke2022grounded} and notebook users more generally~\cite{li2021edassistant}, in that foraging can lead to the discovery of new functionality or ways of doing things.
Among the designs we presented to facilitate this type of task, participants mostly preferred the Code and Output (\figref{fig:disambig}D) for both visualization tasks as well as other situations, noting that it allowed them to \pQ{1}{be your own audit} by checking code and output as task required.
\pX{5} noted that it was valuable because it reminded her of Excel's chart chooser feature.
\pX{10} and \pX{14} worried about the computational resources required to generate alternatives and suggested a hybrid ``\`a la carte'' execution mode in which only selected options were run.
While trust in a system was noted (\pX{2} and \pX{6}) as being mediated by how fast it could supply an answer, \pX{6} and \pX{10} noted that if the system would produce high-quality charts, they would wait multiple minutes.
The Token-alternatives design from \citet{weisz2021perfection} was seen as a valuable mechanism for disambiguation. \pX{12} noted that it allowed them to apply \pQalt{minimum mental effort at learning to make adjustments} to the generated code, and \pX{5} added that they are
\pQalt{super useful if I wanted to explore things.}
\pX{11} suggested that it would be helpful to have documentation integrated with the alternatives.
This suggests that multiple modalities of disambiguation may be valuable for both acceleration tasks (as in the relatively lightweight Token-alternatives) as well as exploration tasks (as in the heavier weight Code and Output).
\lesson{Providing means of verification integrated into the process of code generation (such as \designSpaceItem{Effect} previews) may shorten the accept-validate-repair sequence~\cite{barke2022grounded}, particularly when tuned to different usage modes.}

\begin{table*}[t]
    \caption{
        Takeaways from our interviews. While some elements are always valuable (\eg{} polite interfaces~\cite{whitworth2005polite}), these findings are only grounded in code assistants for computational notebook-based for data science tasks.
    }
    \small
    \begin{center}
        \begin{tabular}{||r|| p{4.4in} r ||}
            \hline
            \multirow{1}{0.5in}{\vspace{-1.5em}\begin{flushright}\textbf{Politeness}\end{flushright}}
                                                                                                                     & 1. Code assistants should be designed so that they are treated as ghostwriters                                                                         & (\secref{sec:documentation})      \\
                                                                                                                     & 2. Assumption of best practices enforcement can be used to provide opinionated guidance                                                                & (\secref{sec:domain})             \\
                                                                                                                     & 3. Code annotation should be ambient or provide value for active interaction                                                                           & (\secref{sec:control})            \\
            \hline

            \multirow{2}{0.5in}{\vspace{-0.9em}\begin{flushright}\textbf{Notebook Patterns}\end{flushright}}
                                                                                                                     & 4. Surfacing multiple ways to control context is useful---\designSpaceItem{Ambient} alone is likely insufficient                                       & (\secref{sec:context-interfaces}) \\
                                                                                                                     & 5. Expectation can be subverted to useful and surprising effects                                                                                       & (\secref{sec:prominence})         \\
                                                                                                                     & 6. Following familiar notebook UI patterns is important for adoption                                                                                   & (\secref{sec:domain})             \\
            \hline
            \multirow{3}{0.5in}{\vspace{-0.9em}\begin{flushright}\textbf{Code Assistant Patterns}\end{flushright}}                                            &
            7. Integration of search and docs is valuable, but code provenance is not---\designSpaceItem{No History} & (\secref{sec:search})                                                                                                                                                                      \\
                                                                                                                     &
            8. Task (or medium)-specific assistants for data science in notebooks are valuable                       & (\secref{sec:domain})                                                                                                                                                                      \\
                                                                                                                     &
            9. Knowledge of the underlying model changes expectations and required affordances                       & (\secref{sec:knowledge})                                                                                                                                                                   \\
                                                                                                                     & 10. Tools like disambiguation (\eg{} via \designSpaceItem{Code} \emph{and} \designSpaceItem{Effects}) can aid the \emph{specification-refinement} loop & (\secref{sec:verification})       \\
            \hline
        \end{tabular}
    \end{center}
    \label{tab:takeaways}
    \Description{ Table showing the key takeaways from the paper, including a higher-level theme, a short description, and a section reference to where that insight was identified.}
\end{table*}

\section{Discussion}

In this paper, we explored the design space of interfaces for AI-powered code assistants in notebooks. To do so we investigated the design space of this style of tool through an analysis of preexisting systems that support code generation in notebooks.
We sought to understand perceptions of several key elements in this space (context, disambiguation, adaptation, and documentation) for a realistic user population of notebooks.
This led us to conduct a semi-structured interview study with 15 professional data scientists.

Through this work, we delineated guidance for designers of future systems in this space, which we summarize in \autoref{tab:takeaways}.
{Participants were unanimously enthusiastic about adapting \surveySystem{Copilot}-style code assistants to the notebook domain.} While not every design was to every participant's taste, it seems that there is ample space to introduce new valuable designs \emph{in addition to} the \designSpaceItem{Ambient} style of \surveySystem{Copilot}.
We suggest, to this end, that the most fruitful ground in this regard lies in creating systems that are specific to a domain task or those tasks things that users do not view as core to their work.
For instance, writing code that is well formatted or follows best practices, or making visualizations.
Within such tasks, disambiguation seems to be an especially powerful means by which to exert control over notebook-based assistants. This may be because the execution loop in which notebooks operate closely mirrors that of the specification-refinement loop.
We suggest that navigating these elements in a polite~\cite{whitworth2005polite} and non-irritating manner is essential for assistant adoption. \designSpaceItem{Ambient} interfaces
offer a low-friction way to provide recommendations; however, their generality can lead to opaque usage and a lack of domain specificity. If such issues are key, then a more explicit design is preferable.

\subsection{Limitations and Generalizability}
As with any study, our studies have limitations that affect their generalizability.

Our design space and interview studies are limited by their predication on design paradigms used in Jupyter and similar ecosystems.
While these patterns are quite common~\cite{lau2020design} (and have been stable for at least a decade),
we do not utilize their ubiquity and commonalities with other systems to generalize beyond code assistants in notebooks.
While some of our observations may be aligned with other domains (such as assistants in IDEs), we emphasize that our findings are local to notebooks,
as investigation of another interface may have led to different findings.

Our design space was limited by our exclusion of search-based systems. Despite their relevance and long history~\cite{liu2021opportunities}, they fell outside of our focus on systems that generate code. While \citet{Sarkar22WhatIsItLike} found generative models are sometimes viewed as being similar to search, they also found that diverge in key ways, such as how search yields mixed-media, while generation gives fixed-media. Future work should investigate the similarities and differences between expected interface affordances between search and generation, so as to better understand how those interaction forms might be more effectively blended~\cite{xu2022ide}.

The slide-based prototypes advantageously allowed for exploration of many different designs but lacked the concreteness of interactive prototypes, which may have elicited different responses had we used them instead.
For instance,  some participants (\secref{sec:prominence}) discussed how annoying a given feature might be and speculated on how that would affect their usage.
While user perception of utility can be elucidative of their actual behavior (per the Technology Acceptance Model \cite{davis1989perceived}), it is not necessarily reflective of their real-world usage.
We sought to address this issue by requiring that our study population have previously used a code assistant so that they could draw on their real-world experiences to shape their expectations of code assistants.
However this measure can only approximate real-world usage, and future work should study the experience of interacting with similar designs in live implementations.
Similarly, the simplicity of the tasks presented may have limited the types of responses and a more complex task type may have elicited other responses.
While a number of respondents (\eg{} \pX{1}, \pX{4}, \pX{6}, \pX{8}, \pX{9}, \pX{11}, and \pX{14}) reflected on their experiences using AI assistance for complex tasks and how the presented design might interact with those tasks (covering topics such as modeling, data pipeline creation, and other NLP tasks), future work might usefully explore the perceptions of our design space in the context of more nuanced tasks.
Our slides may have caused participants to be overly optimistic about our designs, such as the common out-of-order execution issues found in computational notebooks---although this was not always the case, for instance \pX{14} worried about the bugs that might be introduced by the sandbox in \figref{fig:context-models-sidepanel}.
Further, the consistent order in which we presented designs may have biased participant responses,
however, we believe this effect is limited, given the minimal agreement on preferred features.

While work in notebooks is sometimes seen as synonymous with data science (\pX{13}), this is not always the case. Our interviews focused only on a single type of user which may impede the transferability of our findings to other tasks or categories of work.
Users who mostly complete engineering (a divergence noted by \pX{8}) or creative tasks likely have different usage patterns.
Future work should investigate the beliefs and expectations of other user types.

\subsection{Future work}
Notebooks are but a single medium among a vast array of other areas of application. For instance, how code generation should be adapted to end-user tools, like spreadsheets, is an important open question.
\citet{srinivasa2022gridbook} explored weaving natural language prompts into spreadsheets, finding that they can be successfully used by spreadsheet users, suggesting the value of studying such intersections.
Other important application areas that balance graphical and textual specifications (to different proportions) include creative coding systems~\cite{peppler2005creative}, low-code analytics environments that interweave graphical components with analytics queries (such as visual analytics systems like PowerBI or Ivy~\cite{mcnutt2021integrated}), as well as systems that preference graphical representations, such as block-based systems (\eg{} Scratch~\cite{resnick2009scratch}).

Our interview study considered design probes that investigated a limited cross-section of the large number of possibilities described by our design space. Future work might consider the design permutations implied therein.
This might usefully involve explorations about the role of \designSpaceItem{Nonlinear} input systems, such as the code-generating spreadsheets of \surveySystem{Mito}~\cite{mito}, or \designSpaceItem{bidirectionally} synchronized projectional-editors to handle thorny interface tasks, such as visualization annotation.
Further, it may be advantageous to consider code assistants who only output metadata, such as code explanations.

The space of possible functionality available to assistants in notebooks is vast. The mixture of code, data, and graphics presented interactively offers a rich space of opportunities in which an assistant might be usefully interwoven. As code assistants continue to become more powerful, we highlight that careful adaptation to computing environments is critical, as is paying attention to the space of available design alternatives---particularly concerning context control and disambiguation. Models powering these assistants might usefully be designed with the end-user interface in mind,
and facilitate features like search, context-free suggestions, recommendation explanations, and lightweight configuration so that they might be tuned to task on the fly.

\begin{acks}
    We thank our anonymous reviewers and our study participants. In addition we thank Christian Bird, Thomas Zimmermann, Elsie Lee-Robbins, as well as the Microsoft Research VIDA and EPIC teams as a whole for their useful advice and support.
\end{acks}

\bibliographystyle{ACM-Reference-Format}
\bibliography{metabib}

\appendix
\onecolumn %

\section{Appendix}

\begin{figure}[t]
    \centering
    \includegraphics[width=\linewidth]{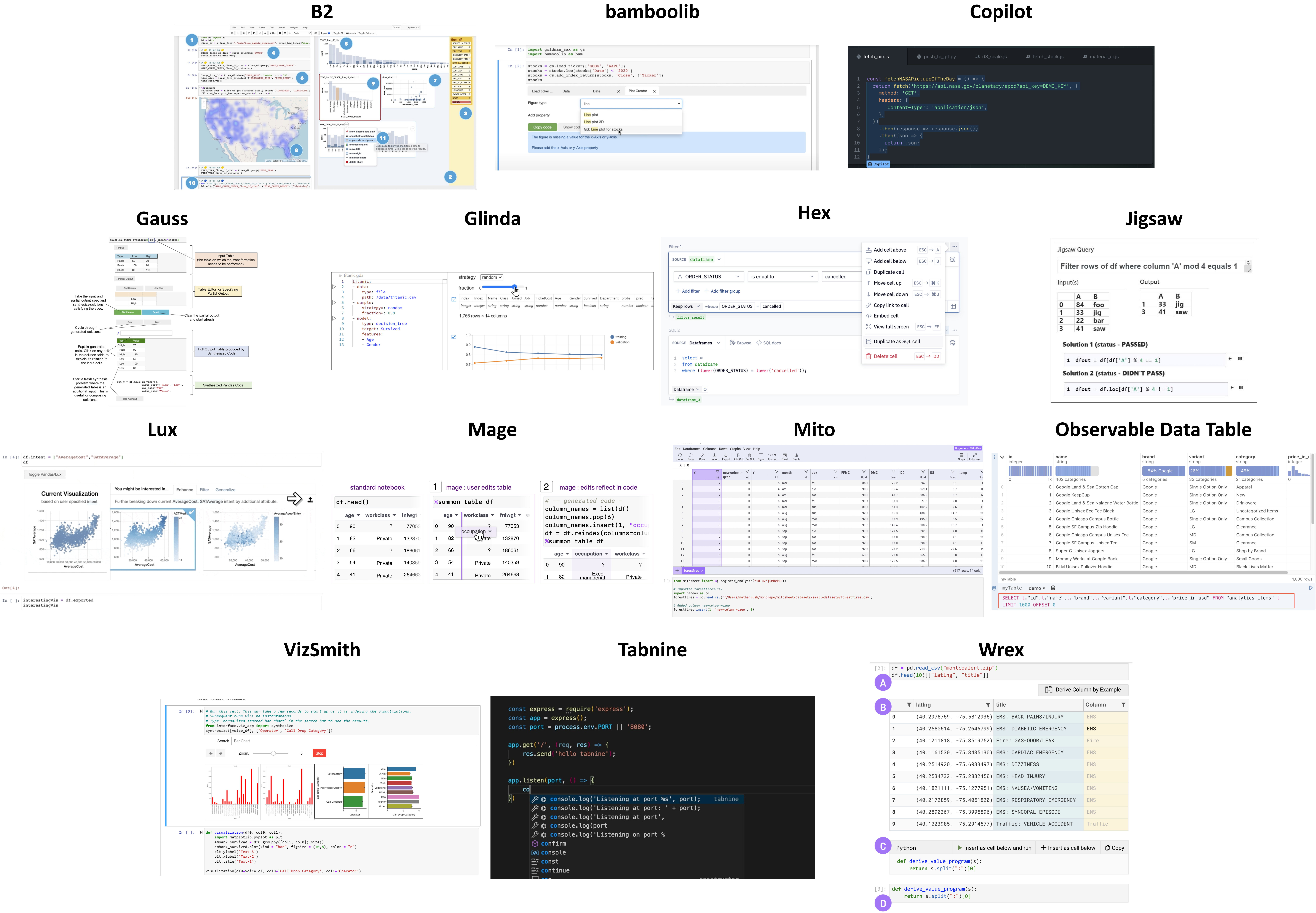}
    \caption{
        A gallery of the systems involved in our design space (\autoref{sec:design-space}). Images are drawn from their relevant papers, with the following exceptions.
        bamboolib: \url{https://www.youtube.com/watch?v=Qni8kX4hSOM},
        hex: \url{https://hex.tech/blog/introducing-no-code-cells/},
        Copilot: \url{https://docs.trymito.io/getting-started/overview-of-the-mitosheet},
        Lux: \url{https://github.com/lux-org/lux},
        Mito: \url{https://docs.trymito.io/getting-started/overview-of-the-mitosheet},
        Observable Data Table: \url{https://observablehq.com/@observablehq/data-table-cell},
        and Tabnine: \url{https://www.youtube.com/watch?v=twPtvZuBrAg}
        The images in this figure are intended to be zoomed into for closer exploration.
        Links accessed 11/11/22.
    }
    \Description{14 screenshots showing images of each of the 14 systems used in the design space survey. They are labeled: B2, bamboolib, Copilot, Gauss, Glinda, Hex, Jigsaw, Lux, Mage, Mito, Observable Data Table, VizSmith, Tabnine, and Wrex.    }
    \label{fig:design-space-pics}
\end{figure}

\begin{figure}[b]
    \centering
    \includegraphics[width=0.9\linewidth]{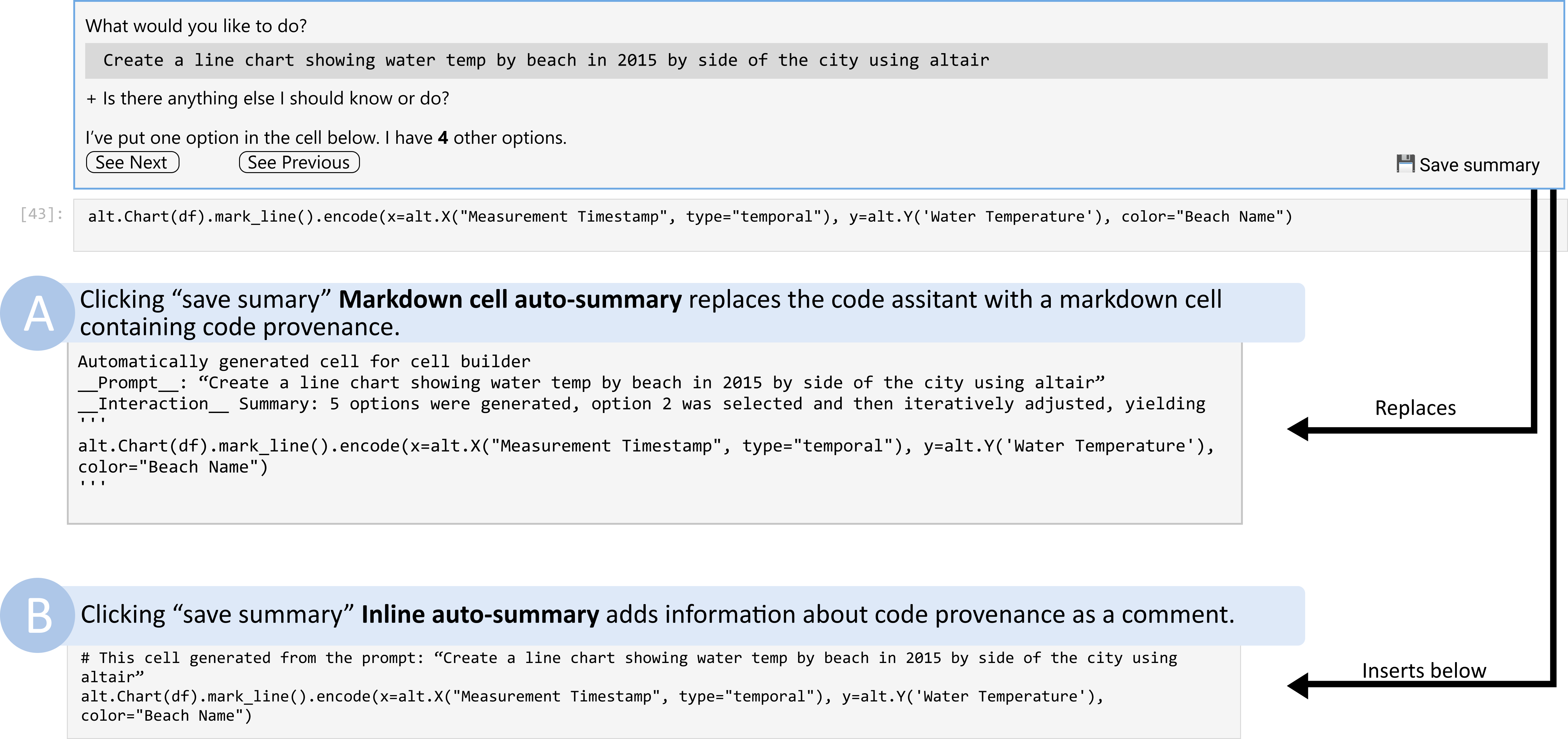}
    \caption{
        Our ``save summary'' button introduces a mechanism to save an automatically generated summary of the interactions with a code assistant that led to a given piece of code.
    }
    \Description{Two screenshots showing proposed documentation features. (A) shows the effect of clicking on a "save summary" button that replaces an inline cell with a markdown cell. It is labeled "Markdown cell auto-summary". (B) shows the effect of clicking on a "save summary" button that inserts information into the generated cell as a comment. It is labeled "Inline auto-summary". }
    \label{fig:documentation-auto-summary}
\end{figure}

\begin{figure}[b]
    \centering
    \includegraphics[width=0.9\linewidth]{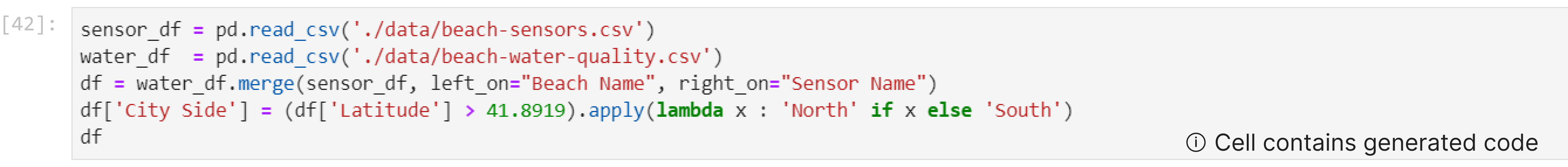}
    \caption{Inline attribution identifies cells that contain code generated by a code assistant and provides information about the interactions that created that code on demand.
    }
    \Description{Screenshot showing a cell with a small marker in the lower right-hand corner that reads "cell contains generated code".}
    \label{fig:documentation-inline-annotation}
\end{figure}

\begin{figure}[t]
    \centering
    \includegraphics[width=0.65\linewidth]{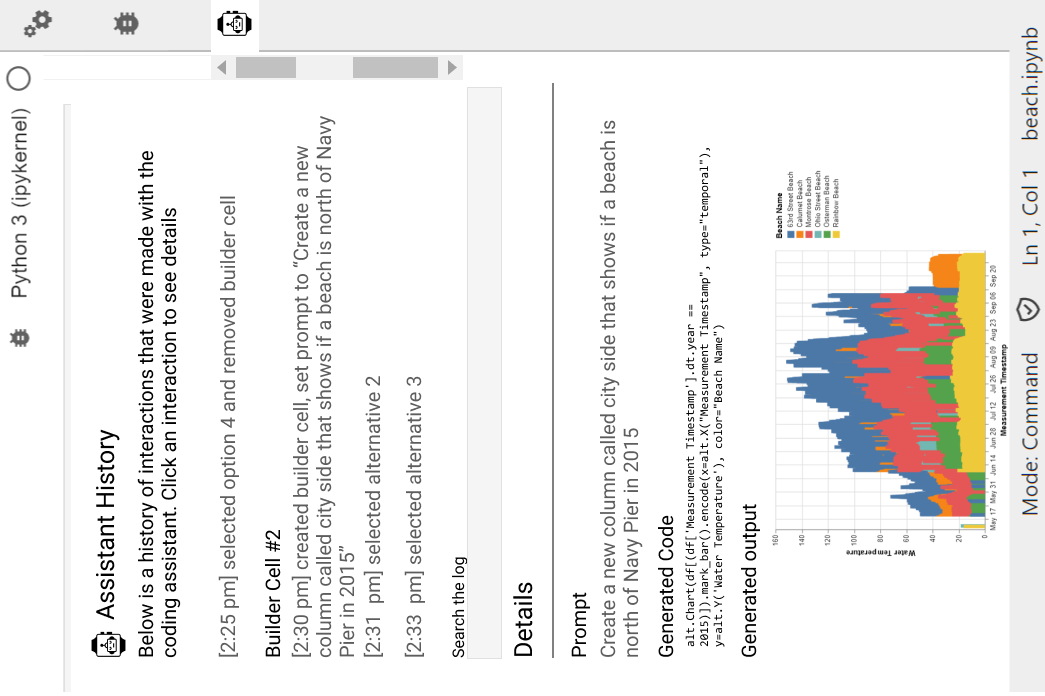}
    \caption{Our explorable log design allowed users to browse a history of their interactions with code assistants in the current notebook.
    }
    \Description{Screenshot showing a Side-panel containing the notebook's history of interactions with code assistants.}
    \label{fig:documentation-history}
\end{figure}

In this appendix, we provide additional figures that we were unable to fit in the main body of the text. In \figref{fig:design-space-pics} we present a gallery of each of the systems used in the design space analysis. In \figref{fig:documentation-auto-summary}, \figref{fig:documentation-inline-annotation}, and \figref{fig:documentation-history} we show annotated versions of the stimuli that were presented to study participants during our interview-design study covering code provenance enhancements.

\end{document}